\documentclass[11pt]{article}
\usepackage[left=3.5cm, right=3.5cm]{geometry}

\usepackage[T1]{fontenc}
\usepackage[utf8]{inputenc}
\usepackage{amstext}
\usepackage{amsfonts}
\usepackage{amsmath}
\usepackage[english]{babel}
\usepackage{graphicx}
\usepackage{booktabs}
\usepackage{comment}
\usepackage{csquotes}
\usepackage[svgnames]{xcolor}
\usepackage{hyperref}
\usepackage{authblk}

\newcommand{\polimi}{Dipartimento di Elettronica, Informazione e Bioingegneria, Politecnico di Milano, Piazza Leonardo da Vinci 32, 20133 Milano, Italy}

\begin{document}
\date{}
\title{A data-driven analysis of the impact of non-compliant individuals on epidemic diffusion in urban settings}

\author{Fabio Mazza, Marco Brambilla, Carlo Piccardi and Francesco Pierri\thanks{francesco.pierri@polimi.it}}
\affil{\polimi}





\maketitle

\begin{abstract}
	Individuals who do not comply with public health safety measures pose a significant challenge to effective epidemic control, as their risky behaviours can undermine public health interventions. 
	This is particularly relevant in urban environments because of their high population density and complex social interactions. 
	In this study, we employ detailed contact networks, built using a data-driven approach, to examine the impact of non-compliant individuals on epidemic dynamics in three major Italian cities: Torino, Milano, and Palermo.
	We use a heterogeneous extension of the Susceptible-Infected-Recovered model that distinguishes between ordinary and non-compliant individuals, who are more infectious and/or more susceptible. 
	By combining electoral data with recent findings on vaccine hesitancy, we obtain spatially heterogeneous distributions of non-compliance.
	Epidemic simulations demonstrate that even a small proportion of non-compliant individuals in the population can substantially increase the number of infections and accelerate the timing of their peak. 
	Furthermore, the impact of non-compliance is greatest when disease transmission rates are moderate.
	Including the heterogeneous, data-driven distribution of non-compliance in the simulations results in infection hotspots forming with varying intensity according to the disease transmission rate. 
	Overall, these findings emphasise the importance of monitoring behavioural compliance and tailoring public health interventions to address localised risks.
\end{abstract}


\section{Introduction}
Over the past decades, considerable research has focused on understanding how individual behaviour influences the dynamics of epidemic outbreaks, demonstrating that behavioural factors can substantially alter the course and severity of disease spread~\cite{poletti_spontaneous_2009, Rizzo_2014,du_how_2021,gozzi_self-initiated_2021,mao_modeling_2014,sontag_misinformation_2022, xia_sir_2012}.
During the COVID-19 pandemic, unsafe health behaviours have been associated with the spread of misinformation on online social networks~\cite{gallotti_assessing_2020,pierri_online_2022,infodemics}.
Health-related non-compliance encompasses a range of risk-enhancing behaviours, including increased vaccine refusal, adoption of unproven or harmful ''treatments'', and the disregard of public health measures such as social distancing and mask mandates~\cite{infodemics,loomba_measuring_2021,tizzani_socioeconomic_2024}.
Vaccine acceptance, in particular, is a topic of great importance for public health~\cite{wal_vaccination_2005,qian_dynamic_2024}. The recent rise in vaccine hesitancy has been linked to the resurgence of vaccine-preventable diseases such as measles, previously believed to be eliminated in many regions~\cite{pananos_critical_2017,pandey_exacerbation_2023}.
Collectively, non-compliant behaviours pose significant challenges during epidemic events, as they may amplify transmission rates and exacerbate the overall impact of outbreaks~\cite{deverna_modeling_2024, baker_epidemic_2021, muntoni_effectiveness_2024}.

To investigate the impact of individuals not complying with public health safety measures on epidemic dynamics, several studies have extended agent-based Susceptible–Infected–Recovered (SIR) models~\cite{kermack_contribution_1927} by allowing disease transmission to depend on individual behavioural status~\cite{mao_modeling_2014,du_how_2021,prandi_effects_2020,gozzi_self-initiated_2021,zheng_interplay_2018}.
In these models, individuals may isolate themselves in response to external mandates (e.g., lockdowns) or perceived risk, thereby effectively removing themselves from the transmission network.
More recently, Deverna et al.~\cite{deverna_modeling_2024} modelled the impact of non-compliant individuals through large-scale data-driven simulations of contact networks combined with measurements of online misinformation consumption on social media. 
Their model increases the infectiousness (i.e., the probability of transmitting the disease) of ''misinformed'' individuals, but keeps susceptibility constant across the population, regardless of behavioural status.
However, earlier research has shown that heterogeneity in both infectivity and susceptibility plays a critical role in shaping epidemic outcomes on contact networks~\cite{miller_epidemic_2007,pastor-satorras_advantage_2022, smir_conf}. 
To fully capture the effects of behavioural differences, models should account for variation in both dimensions.

Urban environments play a critical role in epidemic propagation due to their high population density and central importance to economic and social activities~\cite{dalziel_urbanization_2018,hazarie_interplay_2021,topirceanu_novel_2021}.
Despite this, the specific impact of non-compliant individuals within urban settings remains largely unexplored. 
In particular, few studies have addressed how socio-economic or geographic heterogeneity across city districts may influence epidemic dynamics. 
Understanding these localized differences is essential for designing targeted interventions and improving the effectiveness of public health responses in urban areas.

To explore the impact of non-compliant individuals on epidemic dynamics in urban environments, we construct detailed contact networks that capture in-person interactions among city residents. 
These networks are generated using a data-driven approach that integrates the geographic distribution of the population with age-specific contact matrices derived from survey data~\cite{guarino_inferring_2021}. 
The resulting age-stratified contact graphs assign individuals to localized spatial units (tiles) and reflect realistic patterns of interaction.
We focus on three major Italian cities---Torino, Milano, and Palermo---with populations ranging from approximately 600,000 to 1.2 million residents. 
To capture behavioural heterogeneity, we employ the HeSIR model~\cite{smir_conf}, a heterogeneous extension of the classic SIR framework~\cite{kermack_contribution_1927}, which differentiates between \textit{ordinary individuals} (O) and \textit{non-compliant individuals} (M, short for Misbehaving). 
The latter group represents those who disregard public health recommendations, such as mask-wearing, social distancing, or vaccination. 
Their behaviour increases both their susceptibility to infection and their ability to infect others, making them a key risk factor in disease propagation.
Using efficient epidemic simulation techniques, we quantify how the presence of non-compliant individuals affects key epidemic indicators, examining sensitivity to both model parameters and the proportion of such individuals. 
To incorporate spatial heterogeneity in behaviour, we introduce a tile-level non-compliance propensity score, derived from electoral data and studies linking political orientation to vaccine hesitancy~\cite{pinto_sezioni_2023,paoletti_political_2024}. 
This produces a data-driven spatial distribution of non-compliance within each city.
Our results show that even small proportions of non-compliant individuals can substantially alter epidemic outcomes. 
Moreover, when these individuals are geographically clustered, they can create local infection hotspots that may overwhelm nearby healthcare services. 
These findings underscore the critical importance of monitoring behavioural compliance within urban areas and tailoring public health interventions to address localized risks.

\section{Models and data}

\subsection{Contact graph}\label{sec:graph_model}

In this work, we study the spread of epidemics in urban settings using city-level contact graphs constructed with a data-driven approach, as described next. The resulting contact graphs account both for contacts between family members (\textit{household contacts}) and contacts between friends, colleagues and acquaintances (\textit{social contacts}).

The territory of a city is divided into $N_T$ square tiles and the population of the city is distributed according to high-resolution data obtained from WorldPop~\cite{worldpop}. Each individual is assigned to one of four age cohorts, representing children (under 18 years old), young adults (18 to 34), adults (35 to 64) and elderly (65 and above). The distribution of individuals aross age cohorts is derived from official statistics about the population provided by the Italian National Institute of Statistics (ISTAT,~\cite{istat}).

We can view the contact graph of a city as composed of two subgraphs, one for household contacts and the other for social contacts, according to the approach proposed in~\cite{guarino_inferring_2021} and summarized in Appendix A. The household contact graph contains cliques representing families that reflect general statistics on family composition in terms of age, role and size, e.g., how many adults are parents, how many elderly people live alone and how many children there are per family unit. The social contact graph is constructed using a Fitness-Corrected Block Model (FCBM)~\cite{bernaschi_fitness-corrected_2022}, in which each block contains individuals of a specific age cohort residing in a specific tile. The model assumes that the frequency of interactions among individuals depends on the age cohorts (as discussed, for example, in~\cite{mossong_social_2008}), and decreases with respect to the geographic distance. Moreover, the variability in the number of contacts of an individual can be captured by a lognormally distributed fitness parameter. These assumptions define the expected number of edges between two blocks in the FCBM.

\begin{figure}[!t]
	\centering
	\includegraphics[width=0.8\linewidth]{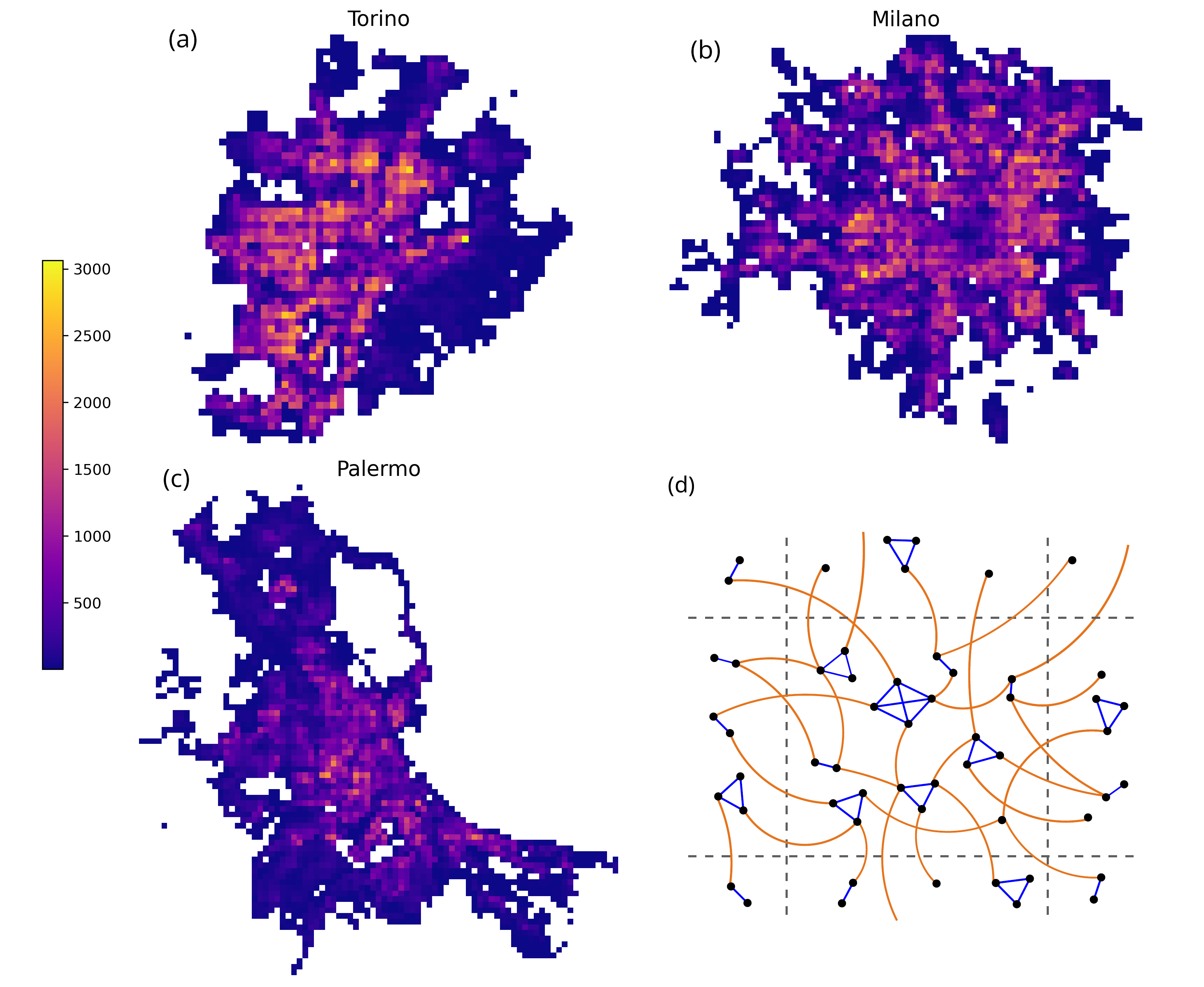}
	\caption{(a-c)~Distribution of the population in tiles for the cities under analysis: each tile is colored according to the number of individuals (see color bar on the left). (d)~Schematic representation of a tile, with household contacts in blue and social contacts in orange – some of the latter connect to tiles not visible in the diagram. The dashed gray lines indicate the boundaries of the tile. 
	}
	\label{fig:popdistrib}
\end{figure}

We constructed the contact graph for three Italian cities, namely Torino, Milano and Palermo. They were chosen because they are among the largest and most populated cities in Italy, represent both Northern and Southern regions and are covered by our data sources. We excluded the capital, Rome, due to its very high population (>2 million) and its geographical extension (>1200 km$^2$), which make the computations very challenging. As shown in Table \ref{tab:networks}, these cities have a population between 630 k and 1.2 M individuals, with the most central tiles having higher density (see Figure \ref{fig:popdistrib}). The graphs contain between 260 k and 529 k households, with nodes having on average 11 edges. More details on the construction of the graphs and city statistics are available in Appendix A.

\begin{table}[!t]
	\centering
	\begin{tabular}{llrrrrrrr}
		\toprule
		City & Tile size & $N$ & $N_T$ & $N_{H}$ & $\left< k \right> \pm \sigma_k$ &  $\left< N_i \right>\pm \sigma_{N_i}$ \\
		\midrule
		Torino & 250m & 878,625 & 1786 & 357,425 & $11.1 \pm 7.3$ & $492\pm 590$ \\
		Milano & 250m & 1,229,360 & 2402 & 529,823 & $11.2 \pm 7.6$ & $512\pm 523$ \\
		Palermo & 200m & 630,242 & 2541 & 260,214 & $11.0 \pm 8.1$ & $248 \pm 318$ \\
		\bottomrule
	\end{tabular}
	\caption{Summary statistics of the contact graphs generated for the cities under analysis. $N$ denotes the number of individuals; $N_T$ is the number of spatial tiles; and $N_H$ is the number of households. The average degree of the graph is given by $\left< k \right>$, and the average number of individuals per tile by $\left< N_i \right>$. The standard deviations of the degree and the number of individuals per tile are denoted by $\sigma_k$ and $\sigma_{N_i}$, respectively.}\label{tab:networks}
\end{table}

\subsection{HeSIR epidemic model}

To study the effect of individuals who do not comply with public health recommendations, hereby non-compliant individuals, we employ the \textit{HeSIR} (\textit{Heterogeneous SIR}) epidemic model, a modified version of the traditional Susceptible-Infectious-Recovered (SIR) model~\cite{kermack_contribution_1927}. This model was proposed and studied in Ref.~\cite{smir_conf}, where the epidemic threshold was derived analytically for a population with homogeneous mixing and for the case of a network with homogeneous contacts. In this section, we introduce the model in detail and discuss its differences from the classical SIR.

\begin{figure}[!t]
	\centering
	\includegraphics[width=0.8\linewidth]{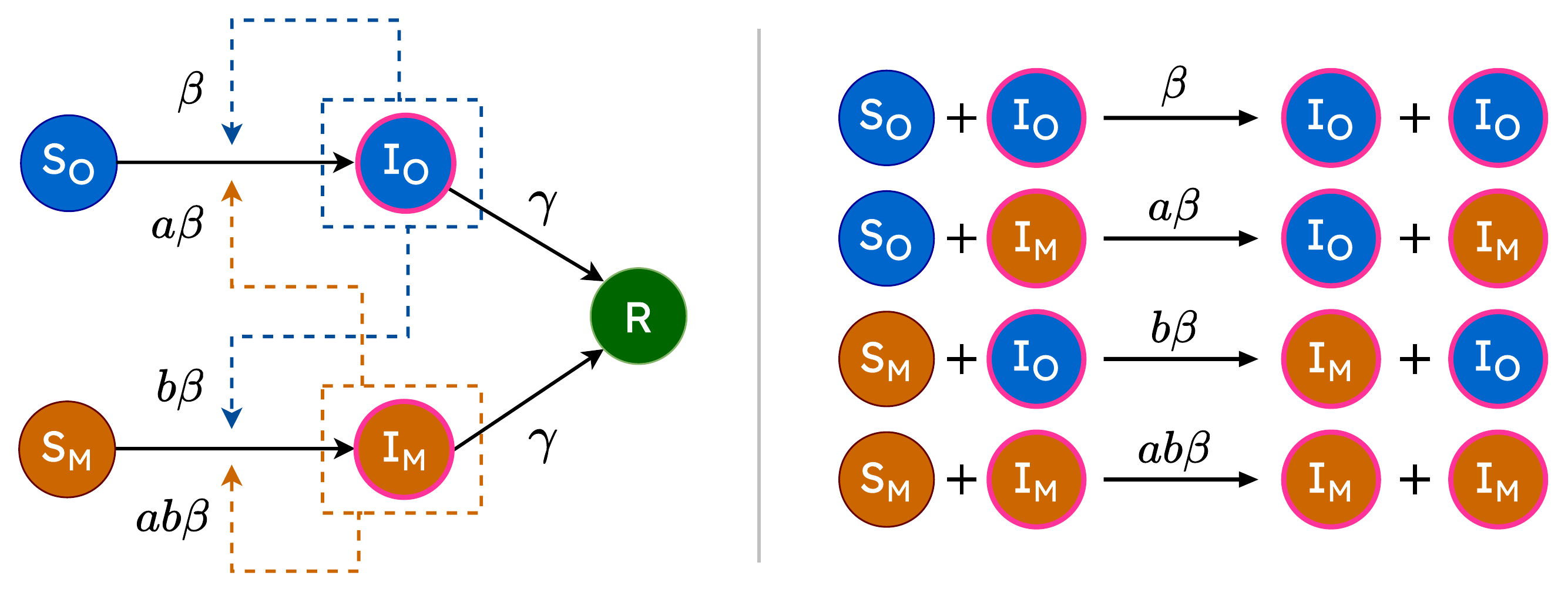}
	\caption{Flow diagram of the HeSIR process. The transition rate from Susceptible to Infectious depends on the classes of both the infector and the infected (with $\beta$ as the baseline transmission rate, and $a, b \geq 1$ representing increased infectivity and susceptibility, respectively). The recovery rate $\gamma$ is uniform across all individuals.
	}
	\label{fig:epimodel}
\end{figure}

The HeSIR model includes the presence of \textit{Non-compliant individuals} (M, short for Misbehaving), who are more likely to infect and be infected than the rest of the population, represented by \textit{Ordinary} individuals (O), who are assumed to be more adherent with public health guidelines. 
More specifically, interactions that occur exclusively between O individuals give rise to infection with rate $\beta$. 
When interactions involve an M individual, the infection rate is multiplied by $a\ge 1$ if the infector belongs to the M class, and by $b\ge 1$ if the susceptible individual is from the M class. 
Thus, the overall rate of disease transmission from an infectious to a susceptible can have four different values depending on the classes M or O of the infector and the susceptible, which are shown in Figure~\ref{fig:epimodel}.
Each individual is assigned to either class O or M \emph{a priori}, i.e.\ before the start of the epidemic process, based on criteria that are detailed below. Class membership remains fixed throughout the epidemic.  
Recovery dynamics follow the standard SIR model: infectious individuals recover at rate $\gamma$, independent of their class (O or M).
When $a=b=1$, the HeSIR model reduces to the usual SIR, as there are no differences between the individuals of the two classes.

In the following, we refer to $\beta$ as the \emph{disease transmission rate}, while $a$ and $b$ denote the \emph{extra infectivity} and \emph{extra susceptibility} of M individuals, respectively.  
This convention, consistent with that adopted in~\cite{miller_epidemic_2007}, emphasizes the microscopic distinction between $a$ and $b$ in our agent-based model of disease transmission.
We will show that while increasing either $a$ or $b$ results in a higher overall disease transmission rate, there is a microscopic distinction when this model is implemented in an agent-based framework with an underlying contact graph, as in our case.  
When $a = 1$ and $b > 1$, the infection rate is elevated only for M individuals, with O individuals remaining unaffected. In contrast, when $a > 1$ and $b = 1$, M individuals are capable of infecting a larger portion of the population, irrespective of the class of the individuals they contact.

It is worth noting that an alternative, though mathematically equivalent, perspective would be to consider non-compliant behaviour as the default risk level and model compliance as a reduction in infection probability compared to that baseline (e.g.~\cite{du_how_2021, hong_co-evolution_2022}). The rationale of our choice lies in the situation experienced in Italy during the COVID-19 pandemic, with a dominant fraction of compliant individuals (e.g., vaccine coverage was approximately 85\% with two doses toward the end of 2021~\cite{zamagni_covid-19_2022}) which naturally leads to considering non-compliant individuals as exceptions.

We simulate epidemics using the HeSIR model on the urban contact graphs described in the previous section.  
In all simulations, individuals are initially in the susceptible state, except for a small set of randomly selected infected individuals---the ``patient zeros''. 
The initial number of infected individuals is set to $I_0 = 50$ for Torino and Milano, and $I_0 = 30$ for Palermo, which has a smaller population.  
For the assignment of individuals to classes M and O, we consider two alternative approaches:

\begin{itemize}
	\item \textit{Uniform distribution}: each individual is assigned to class M with probability $p_M$, regardless of geographical location;
	\item \textit{Data-driven distribution}: an individual belonging to geographic tile $i$ is assigned to class M with probability $p_{M,i}$. 
	As described in the next section \ref{sec:pvhe_distr}, the values of $p_{M,i}$ are computed based on political voting behaviour in each tile.
\end{itemize}
The variable $p_M$ represents the overall proportion of non-compliant individuals in the population, and we experiment with values in the set $\{0.1, 0.2, 0.3\}$.

Owing to the large population size, simulating the dynamics in continuous time (i.e. by using algorithms like Gillespie~\cite{gillespie_exact_1977}) is computationally prohibitive.  
To address this, we approximate the model dynamics by discretizing time into small intervals of $\delta t = 0.1$, with the recovery rate fixed at $\gamma = 0.1$. The value of $\delta t$ was selected through repeated trials with different discretizations, ensuring that the deviation from the continuous-time dynamics remains minimal.  
This approach enables the simulation of epidemic outbreaks involving more than 100 k individuals with minimal loss of accuracy. A link to the code to reproduce our experiments is available in the Data Availability statement at the end of the manuscript.

\subsection{Data-driven distribution of non-compliant individuals}\label{sec:pvhe_distr}

To obtain a realistic geographical distribution of non-compliant individuals, we adopt a data-driven approach informed by recent research on the relationship between vaccine hesitancy and political affiliation~\cite{barbieri_political_2021, paoletti_political_2024}.  
In recent years, considerable attention has been devoted to the impact of vaccine hesitancy and refusal, particularly during the COVID-19 pandemic, on public health intervention~\cite{cadeddu_beliefs_2020, loomba_measuring_2021, barbieri_political_2021, pierri_online_2022}.  
In particular, Paoletti et al.~\cite{paoletti_political_2024} analyse Twitter conversations from a large number of users across various European countries, assigning each user a \textit{Vaccine Hesitancy Endorsement} (VHE) score.
This score is then regressed against both the fraction of politicians followed in each political party and a set of network features. The outcome is a set of VHE fit coefficients---publicly available from~\cite{paoletti_political_2024}---that quantify vaccine hesitancy as a function of political orientation.

In our work, we make use of these coefficients under the hypothesis that vaccine hesitancy serves as a proxy for individual non-compliance with public health directives.
For each geographic tile $i$ specified in Section~\ref{sec:graph_model}, we define the \emph{proportion of non-compliant individuals}, denoted $r_i$, which is rescaled to the interval $[0,1]$---where $0$ corresponds to minimal non-compliance and $1$ to maximal non-compliance---independently for each city.  
To compute the values of $r_i$ and thereby obtain the spatial distribution of the non-compliant population, we combine the VHE fit coefficients with electoral results.  

\begin{figure}[!t]
	\centering
	\includegraphics[width=1.\linewidth]{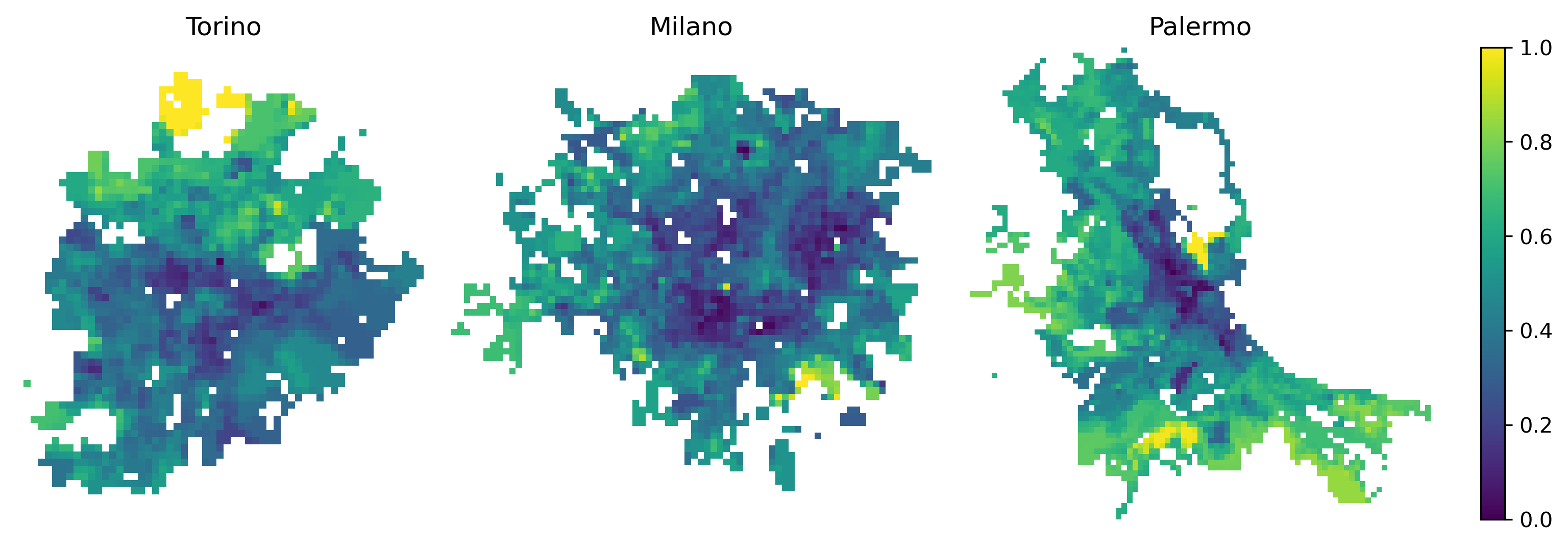}
	\caption{Data-driven distribution of the proportion of non-compliant individuals, $r_i$, scaled such that $0$ represents the minimum level and $1$ the maximum level of non-compliance. Darker colors correspond to smaller values, using the same color normalization for the three cities.
	}
	\label{fig:pVHEs}
\end{figure}

Election data are originally reported at the level of electoral districts, which vary in size and shape. To obtain a measure of non-compliance for the tiles, votes are projected proportionally to their area of overlap. Mathematically, this is expressed as  
\begin{equation}
	x_{i} = \sum_{k} \frac{A_{ki}}{A_i} \sum_{p} \beta_p v_{pk},
\end{equation}  
where $x_i$ is the proportion of non-compliant individuals in tile $i$ before rescaling (see eq.~(\ref{eq:r_i}) below), $A_i$ is the area of tile $i$, $A_{ki}$ the overlap area between voting district $k$ and tile $i$, $\beta_p$ the VHE fit coefficient for party $p$, and $v_{pk}$ the fraction of votes received by party $p$ in district $k$.  
The vote counts and geographical boundaries are obtained from~\cite{pinto_sezioni_2023}; specifically, we use the results from the 2022 Italian general election.
Finally, we rescale $x_i$ in $[0, 1]$ by letting
\begin{equation}\label{eq:r_i}
	r_i = \frac{x_i - \min{x_i}}{\max{x_i}-\min{x_i}}.
\end{equation}
Figure~\ref{fig:pVHEs} shows the distribution of $r_i$ in Torino, Milano, and Palermo. In all cities, there is an evident low concentration of non-compliant individuals in the city center, whereas some peripheral areas exhibit localized clusters with high proportions of non-compliance.

From the proportion of non-compliant individuals $r_i$, we define the non-compliance probability $p_{M,i}$ for tile $i$, representing the probability that an individual in tile $i$ is non-compliant (class M). 
Given a target overall fraction $p_M$ of non-compliant individuals in the city, we impose the constraint that the expected total number of non-compliant individuals satisfies 
\[
\sum_i p_{M,i} N_i = p_M N,
\]  
where $N_i$ is the population of tile $i$ and $N = \sum_i N_i$ is the total population. Assuming $p_{M,i}$ is proportional to $r_i$, the constraint leads to  
\begin{equation}
	p_{M,i} = \frac{r_i}{\langle r \rangle} p_M,
\end{equation}  
where  
\[
\langle r \rangle = \frac{1}{N} \sum_i r_i N_i,
\]  
ensuring that the total expected number of non-compliant individuals matches the target $p_M N$.

In summary, unlike the \textit{uniform distribution}, where every individual in the city has the same probability $p_M$ of being non-compliant (class M), the \textit{data-driven distribution} uses $p_M$ as the average fraction of non-compliant individuals city-wide. 
In this case, the probability that an individual in tile $i$ is non-compliant is given by $p_{M,i}$, which varies spatially according to the local characteristics.


\section{Results}

\subsection{The impact of non-compliant individuals: SIR vs HeSIR with uniform distribution}

We analyze the impact on epidemic spreading of introducing non-compliant (M) individuals into the population, assuming in this section that they are uniformly distributed with probability $p_M$. 
To cope with the stochastic nature of the process, results are averaged over 100 simulations for each parameter set; in each simulation, the M individuals and the initial infected (``patient zeros'') are randomly selected. We present here only the results for Milano, as those for the other two cities are qualitatively similar, with minor differences discussed later. 
Results for the other cities are in the Supplementary Material file.

Figure~\ref{fig:AR_global_new} (top panels) shows that when $a,b > 1$---representing a scenario where non-compliant individuals are both more infectious and more susceptible than the rest of the population---the total number of infected individuals increases significantly compared to the baseline SIR model, for all values of $\beta$.
These effects are quantified by the Attack Rate (AR), $R_{\infty}$, defined as the total number of individuals recovered at the end of the outbreak.
We also observe that below a certain threshold value of $\beta$, the epidemic fails to spread, resulting in $R_{\infty} \approx 0$. 
This behaviour is a well-known characteristic of the SIR model. 
Notably, the critical transmission rate $\beta_c$ required for an outbreak decreases as $a$ and $b$ increase, and to a lesser extent as the fraction of non-compliant individuals $p_M$ increases. 
In other words, the presence of non-compliant individuals lowers the transmission rate threshold $\beta$ needed for the disease to infect a significant portion of the population, especially when non-compliant individuals exhibit substantially higher infectivity and susceptibility (larger $a$ and $b$).

The bottom panels of Figure~\ref{fig:AR_global_new} display the difference in Attack Rate (AR) between the HeSIR model with a uniform distribution of non-compliant individuals and the baseline SIR model.
The increase in AR is striking, peaking around $\beta \langle k \rangle = 0.07$, where an additional 40\% of the population becomes infected for $p_M = 0.3$. 
This peak aligns with the epidemic threshold of the SIR model, at which $R_{\infty} \approx 0$. 
However, the AR difference remains substantial even at intermediate values of $\beta$. 
For example, at $\beta \langle k \rangle = 0.1$, the additional fraction of infected individuals ranges from 5\% to 15\% for $p_M = 0.1$, from approximately 10\% to 25\% for $p_M = 0.2$, and from about 12\% to 33\% for $p_M = 0.3$, with the highest values corresponding to the riskiest behaviours (larger $a,b$). 
Remarkably, the extra AR decreases rapidly at very high transmission rates, indicating that the presence of non-compliant individuals has a more pronounced impact when disease transmissibility is moderate.

Further examination of the bottom panels in Figure~\ref{fig:AR_global_new} reveals a nuanced interaction between the parameters of extra infectivity $a$ and extra susceptibility $b$  for non-compliant individuals. 
Comparing the cases $a=1,b=2$ (blue), $a=1,b=3$ (orange), and $a=2,b=1$ (green), we observe that at low $\beta$ values, the blue and green curves are close and share a similar epidemic threshold.
However, as $\beta$ increases, the green curve rises above the blue and approaches the orange, eventually overlapping it across all $p_M$ values.
This suggests that increased infectivity alone ($a=2,b=1$) leads to more infections than increased susceptibility alone ($a=1,b=2$), while both yield the same threshold. 
Moreover, raising susceptibility only ($a=1,b=3$) results in a higher AR and a lower threshold at low transmission rates, but at higher $\beta$ values, high susceptibility produces attack rates comparable to those from increased infectivity ($a=2,b=1$).

\begin{figure}[!t]
	\begin{centering}
		\includegraphics[width=0.9\linewidth]{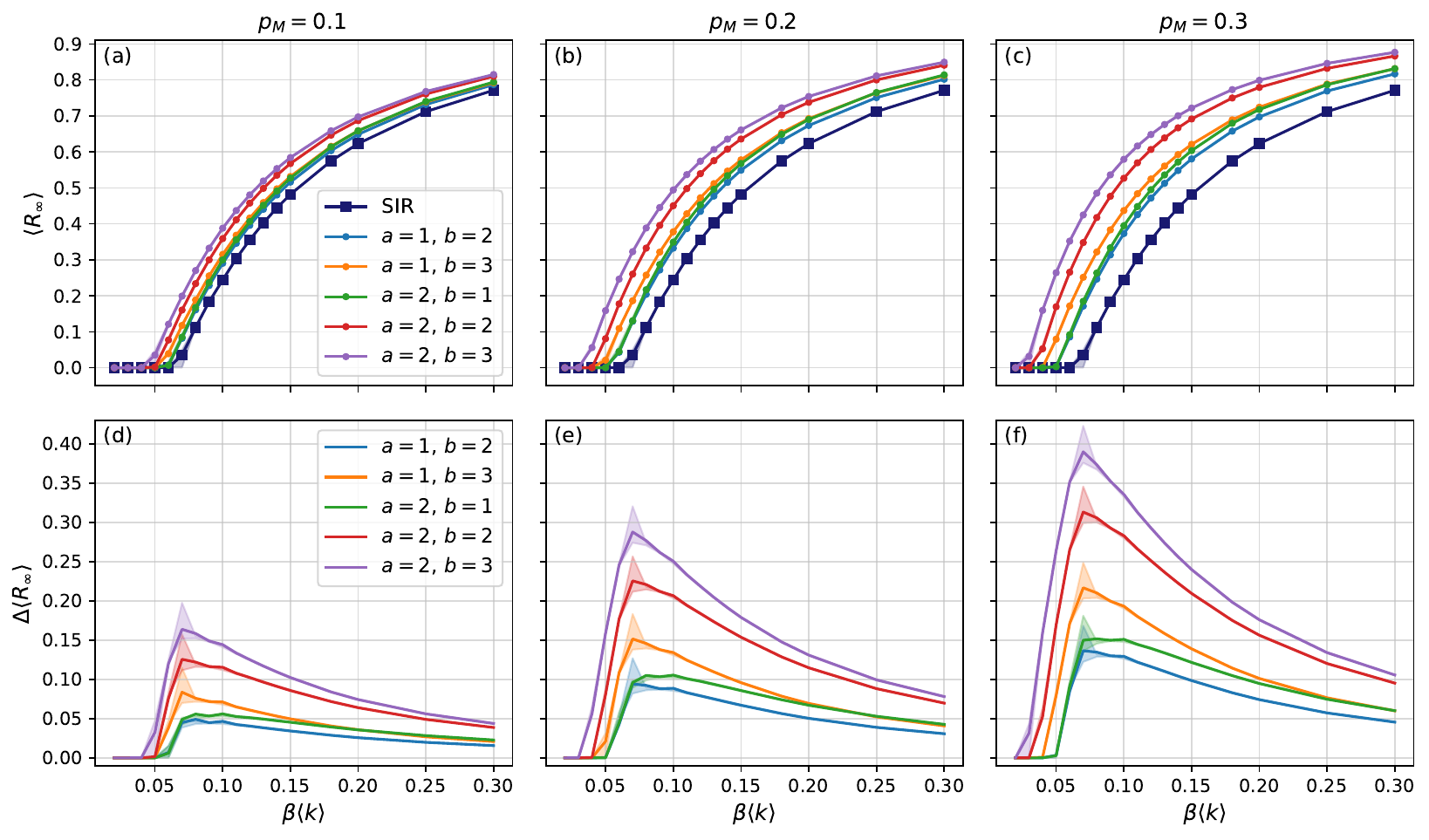}
		\par\end{centering}
	\caption{Simulations of the HeSIR model in the city of Milano with a uniform distribution of non-compliant individuals. Panels (a-c): total fraction of infected individuals (Attack Rate) as a function of the product of transmission rate $\beta$ and average graph degree $\left< k \right>$, for varying values of $a$, $b$, and $p_M$. Panels (d-f): difference in Attack Rates between the HeSIR and baseline SIR models, with the same $p_M$ as the panel above. In all panels, lines indicate the mean and shaded areas correspond to inter-quartile range.
	}\label{fig:AR_global_new}
\end{figure}

\begin{figure}[!t]
	\centering
	\includegraphics[width=0.95\columnwidth]{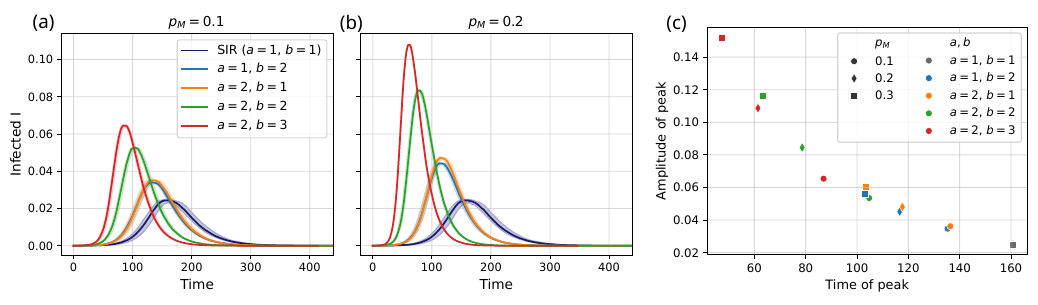}
	\caption{Epidemic outbreak simulations of the HeSIR model in the city of Milano with a uniform distribution of non-compliant individuals ($\beta \langle k \rangle = 0.1$). Panels (a-b): time evolution of the infected population fraction for two values of the non-compliant fraction $p_M$.  Lines indicate median values and shaded areas the interquartile range. Panel (c): scatter plot of epidemic peak time versus peak amplitude, averaged over 100 simulations, for various combinations of model parameters. Points located in the upper-left region indicate more severe outbreaks (earlier and higher peaks). The baseline SIR scenario, with no behavioural heterogeneity ($a = b = 1$), is shown in grey.}
	\label{fig:Global_dynamics}
\end{figure}

We conclude by comparing the dynamic properties of the HeSIR and SIR models, focusing on the timing and magnitude of the epidemic peak, defined as the maximum number of infected individuals. 
Figure~\ref{fig:Global_dynamics} shows that increasing the fraction of non-compliant individuals ($p_M$) and/or decreasing the level of compliance with health measures (higher values of $a$ and $b$) accelerates the epidemic dynamics, causing the peak to occur earlier and with greater intensity.
This faster disease spread poses serious challenges for public health management, as it leads to increased strain on healthcare resources, including hospitals and emergency services. 
The right panel of Figure~\ref{fig:Global_dynamics} compares the peak timing and amplitude across different parameter sets, revealing that various combinations of $a$, $b$, and $p_M$ can produce similar epidemic peaks. 
For instance, just 10\% ``strongly'' non-compliant individuals ($a=2,b=2$) can trigger an outbreak as severe as one caused by 30\% of ``mildly'' non-compliant individuals ($a=1,b=2$).

\subsection{The impact of non-compliant individuals: HeSIR with uniform vs data-driven distribution}

We now assess how the outcomes of epidemic simulations change when non-compliant individuals are assigned using the data-driven distribution rather than uniformly. As detailed in Sec.~\ref{sec:pvhe_distr}, the overall fraction of non-compliant individuals is fixed by the parameter $p_M$. In the data-driven case, however, the individual probability $p_{M,i}$ of belonging to the non-compliant class in tile $i$ is proportional to the tile-specific proportion $r_i$, with $p_M$ corresponding to the weighted average of $p_{M,i}$ across tiles, using the tile populations $N_i$ as weights. To enable a consistent comparison, all results presented in this section are obtained using the same value of $p_M$ for both the uniform and data-driven distributions.

\begin{figure}[!t]
	\centering
	\includegraphics[width=1.0\linewidth]{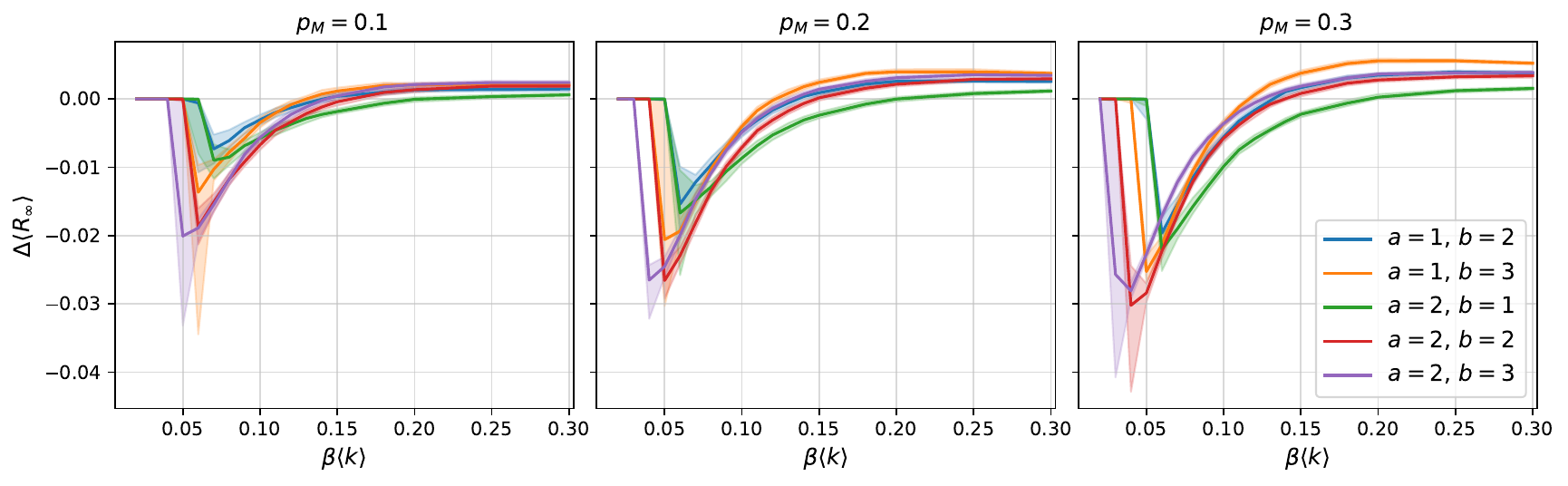}
	\caption{Difference in Attack Rate in the city of Milan between the data-driven and uniform distributions of non-compliant individuals, for different fraction of non-compliant individuals $p_M$. Negative values indicate fewer infections under the data-driven distribution, and positive values indicate more. Lines show median values and shaded areas the interquartile range from the simulations.}
	\label{fig:diff_AR_datadriven}
\end{figure}

First, we observe that the data-driven distribution has a limited effect on the aggregate epidemic dynamics at the city scale. 
As shown in Figure~\ref{fig:diff_AR_datadriven}, the difference in AR compared to the uniform distribution remains below 3\% of the total population. 
While this still corresponds to a substantial absolute number of individuals in large urban areas, the effect is largely confined to low values of the transmission rate $\beta$, it exhibits considerable variability across simulation runs, and it is not consistently observed across all cities (see Supplementary Material for further details).

\begin{figure}[!t]
\centering
\includegraphics[width=0.9\columnwidth]{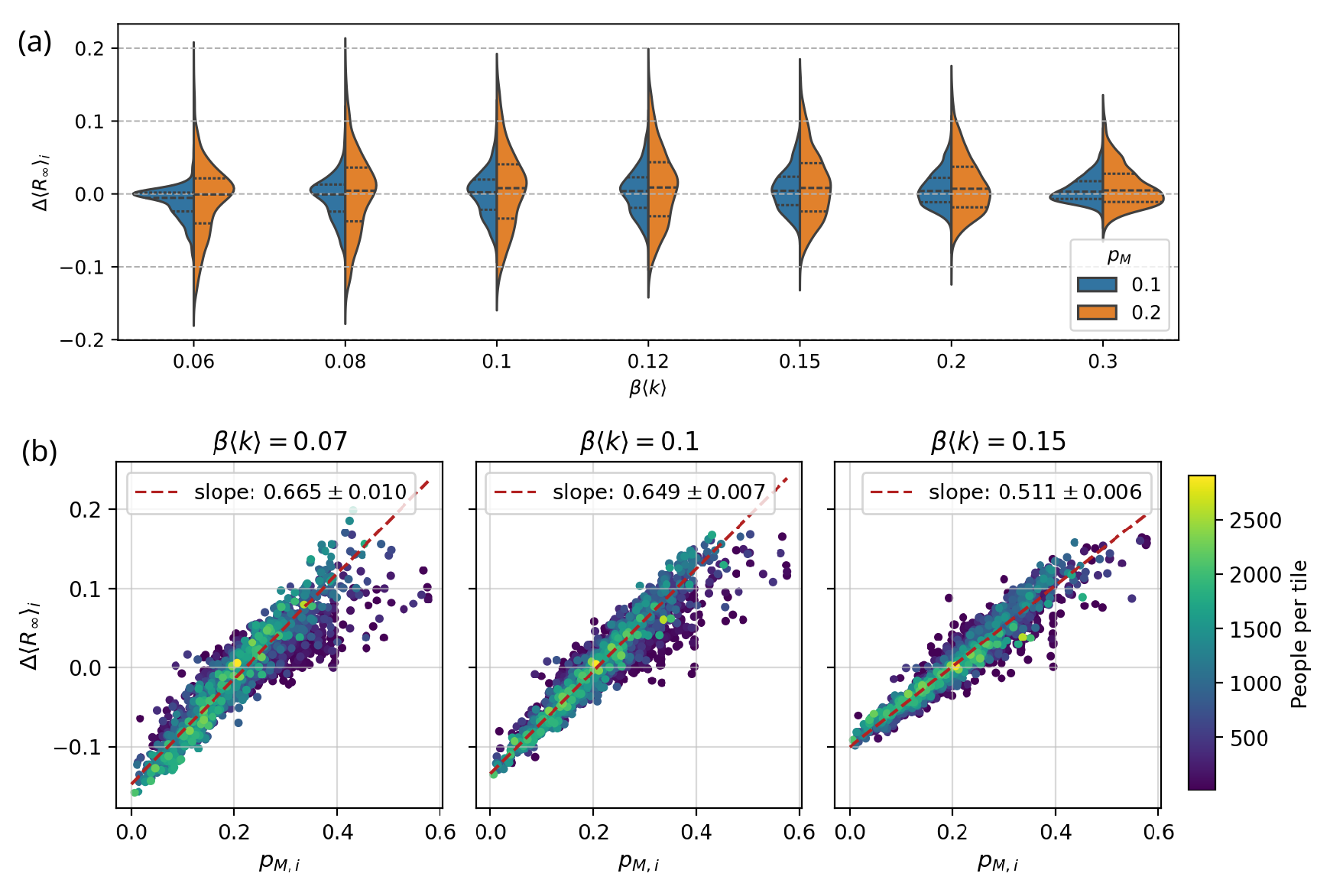}
\caption{Simulations with the HeSIR model in the city of Milano. Panel (a): Distribution of the difference $\Delta \left< R_{\infty}\right>_i$ in the AR of each tile $i$ between the data-driven and the uniform case, for two values of $p_M$ and for different transmission rates $\beta$. Panel (b): the scatter plots show the dependence of $\Delta \left< R_{\infty}\right>_i$ on the fraction $p_{M,i}$ of non-compliant individuals (the city average is $p_M=0.2$). Each dot corresponds to a tile, whose population is color-coded (see colorbar on the right). The red dashed line is the best linear fit (weighted by the tile population): the slope value (with 95\% CI) is reported at the top of the panel. The other simulation parameters are set to $a=2$, $b=3$.}\label{fig:DeltaAR_tiles}
\end{figure}

Nonetheless, the complexity of the epidemic dynamics becomes more evident when examined at the local (i.e., tile) level. 
We begin by analyzing the difference $\Delta \left< R_{\infty}\right>_i$, namely the change in the average AR of tile $i$ between the data-driven and uniform scenarios, computed across simulation runs. 
Figure~\ref{fig:DeltaAR_tiles} (top panel) shows the distribution of $\Delta \left< R_{\infty}\right>_i$ for two values of $p_M$ and several values of the transmission rate $\beta$. 
While the median differences remain close to zero, consistent with the aggregate results discussed above, the distributions exhibit pronounced tails reaching up to approximately 0.2, indicating that in some tiles, the data-driven assignment leads to an excess of infected individuals amounting to 20\% of the tile's population. 
As $\beta$ increases, the distribution becomes narrower, suggesting a homogenization of epidemic outcomes at higher transmission rates. 
This local variability is closely related to spatial heterogeneity in the fraction of non-compliant individuals. Indeed, we find a strong correlation between $\Delta \left< R_{\infty}\right>_i$ and the local fraction $p_{M,i}$ of non-compliant individuals, as illustrated in Figure~\ref{fig:DeltaAR_tiles} (bottom panels): beyond statistical noise, tiles with $p_{M,i} > p_M$ tend to show a positive AR difference, while those with $p_{M,i} < p_M$ tend to show a negative one.

To quantify the relationship between the local fraction of non-compliant individuals $p_{M,i}$ and the resulting change in attack rate, we perform a linear regression of $\Delta \left< R_{\infty} \right>_i$ against $p_{M,i}$, weighting each tile by its population.
The fitted lines for selected scenarios are shown in the bottom panels of Fig.~\ref{fig:DeltaAR_tiles}. 
We observe that the slope of the fit, which captures the sensitivity of the local outbreak severity to variations in non-compliance, varies markedly with the transmission rate $\beta$. To explore this dependence systematically, we compute the slope of the linear fit across a wide range of parameter combinations $(\beta, a, b)$, with results summarized in Fig.~\ref{fig:tiles_fits} (left panel). 
Remarkably, all cities exhibit consistent qualitative trends (see Supplemental Material): the slope initially increases with $\beta$, reaches a peak, and subsequently declines toward an asymptotic value.
This decline reflects the diminishing marginal effect of behavioural heterogeneity as disease transmissibility increases. 
When $\beta$ is high, even modest levels of non-compliance suffice to drive widespread infection, reducing the impact of spatial clustering of non-compliant individuals.

These findings indicate the existence of an intermediate regime of transmissibility in which the spatial distribution of non-compliance exerts maximal influence on local epidemic dynamics. 
Furthermore, we find that increasing either the relative infectivity $a$ or susceptibility $b$ of non-compliant individuals consistently amplifies the slope, strengthening the dependence of local attack rates on $p_{M,i}$.

We next examine the amplitude of the epidemic peak at the local scale, focusing on the excess peak prevalence $\Delta I^{\text{peak}}_i$, defined as the difference in the maximum number of infected individuals in tile $i$ between the data-driven and uniform distributions, averaged across the epidemic simulations.
As with attack rates, we observe a strong correlation between $\Delta I^{\text{peak}}_i$ and the local proportion of non-compliant individuals $p_{M,i}$: specifically, tiles where $p_{M,i} > p_M$ tend to exhibit positive peak differences, while those with $p_{M,i} < p_M$ show negative values.
To quantify this relationship, we perform a linear regression of $\Delta I^{\text{peak}}_i$ against $p_{M,i}$, again weighting by tile population. 
The results, displayed in the right panel of Fig.~\ref{fig:tiles_fits}, show that the slope of this relationship increases with the transmission rate $\beta$, but, unlike the case of attack rates, eventually stabilizes at a plateau rather than decreasing. 
This suggests that the local impact of non-compliance on peak intensity remains substantial even in high-transmission regimes.
Consistent with earlier findings, the magnitude of this dependence grows with the parameters $a$ and $b$, confirming that both increased infectivity and susceptibility among non-compliant individuals exacerbate the heterogeneity of epidemic dynamics across space.

Figure~\ref{fig:tiles_fits} also illustrates how the slopes of the linear fits vary with different values of the extra infectivity $a$ and extra susceptibility $b$ of non-compliant individuals. 
Notably, the slopes are consistently lower for the configuration $a=2, b=1$ compared to $a=1, b=2$. This suggests that increased susceptibility ($b>1$) plays a more critical role than increased infectivity in generating local variations in epidemic outcomes, as it results in a greater number of infections in tiles with higher concentrations of non-compliant individuals.
When both $a$ and $b$ are increased simultaneously, the two effects combine, amplifying the dependence of local epidemic outcomes on the spatial distribution of non-compliance. 
This is particularly evident when comparing the slopes for the configuration $a=2, b=2$ against those for $a=1, b=2$ or $a=2, b=1$: the former yields slopes that are nearly double in the case of peak prevalence differences ($\Delta I^{\text{peak}}_i$) and almost twice as large for attack rate differences ($\Delta \langle R_\infty \rangle_i$), highlighting the synergistic effect of increased infectivity and susceptibility.

\begin{figure}[!t]
\begin{centering}
	\includegraphics[height=0.3\columnwidth]{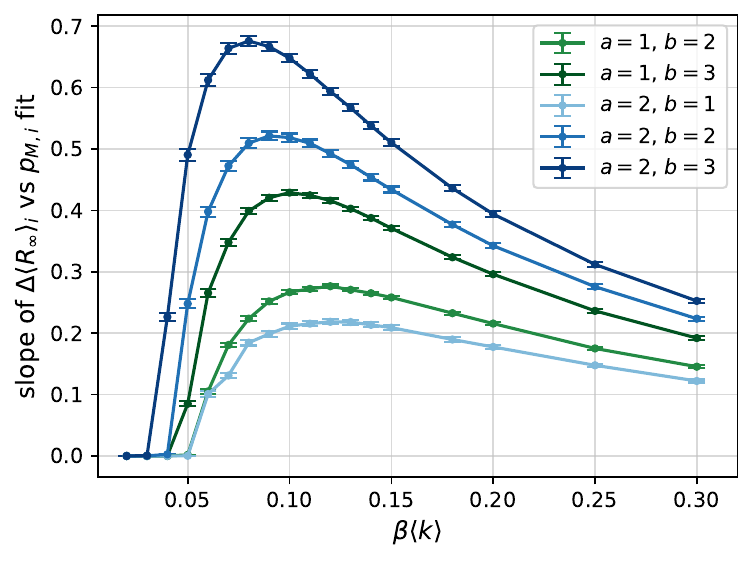}
	\includegraphics[height=0.3\columnwidth]{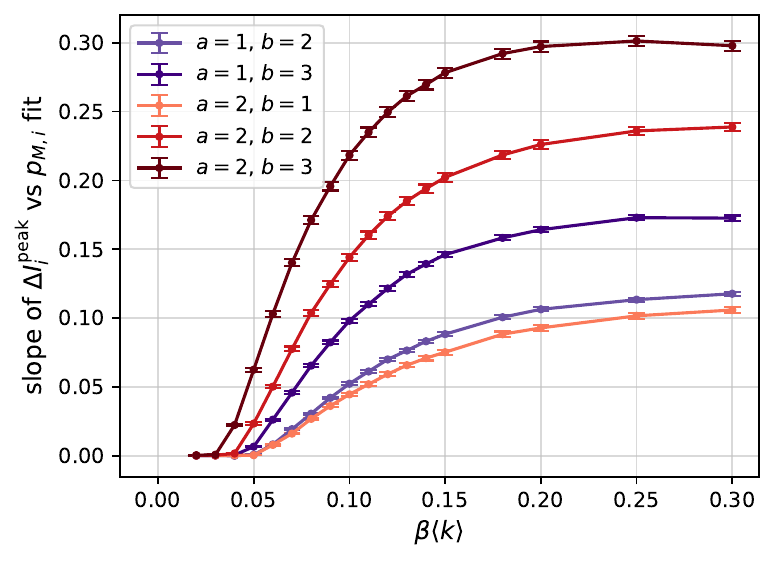}
	\par\end{centering}
\caption{Simulations with the HeSIR model in the city of Milano using the data-driven distribution. \textbf{Left:} Slope of the best linear fit of $\Delta \left< R_{\infty}\right>_i$ versus $p_{M,i}$ as a function of $\beta$, for various values of $a$ and $b$ (see Figure \ref{fig:DeltaAR_tiles}). \textbf{Right:} Slope of the best linear fit of $\Delta I^{\textrm{peak}}_{i}$ versus $p_{M,i}$ as a function of $\beta$, for different values of $a$ and $b$. Each point corresponds to a simulation set used for fitting; error bars represent the 95\% confidence intervals. The average fraction of non-compliant individuals is fixed at $p_M=0.2$.
}
\label{fig:tiles_fits}
\end{figure}

\subsection{Effects of contact graph randomization}

As detailed in Sec. \ref{sec:graph_model}, the contact graph model employed in this study captures the complexity of both household and social interactions, with social contacts influenced by factors such as age class, geographic distance, and individual fitness. 
To evaluate the impact of these modelling assumptions, we construct a randomized version of the contact graph using the configuration model~\cite{bailey_siam_2018}, which preserves the original degree sequence but randomly reassigns edges. 
This \textit{shuffled contact graph} maintains the total number of connections but effectively eliminates almost all household ties as well as correlations related to age, distance, and individual fitness in social contacts.

\begin{figure}[!t]
\centering
\includegraphics[width=0.8\linewidth]{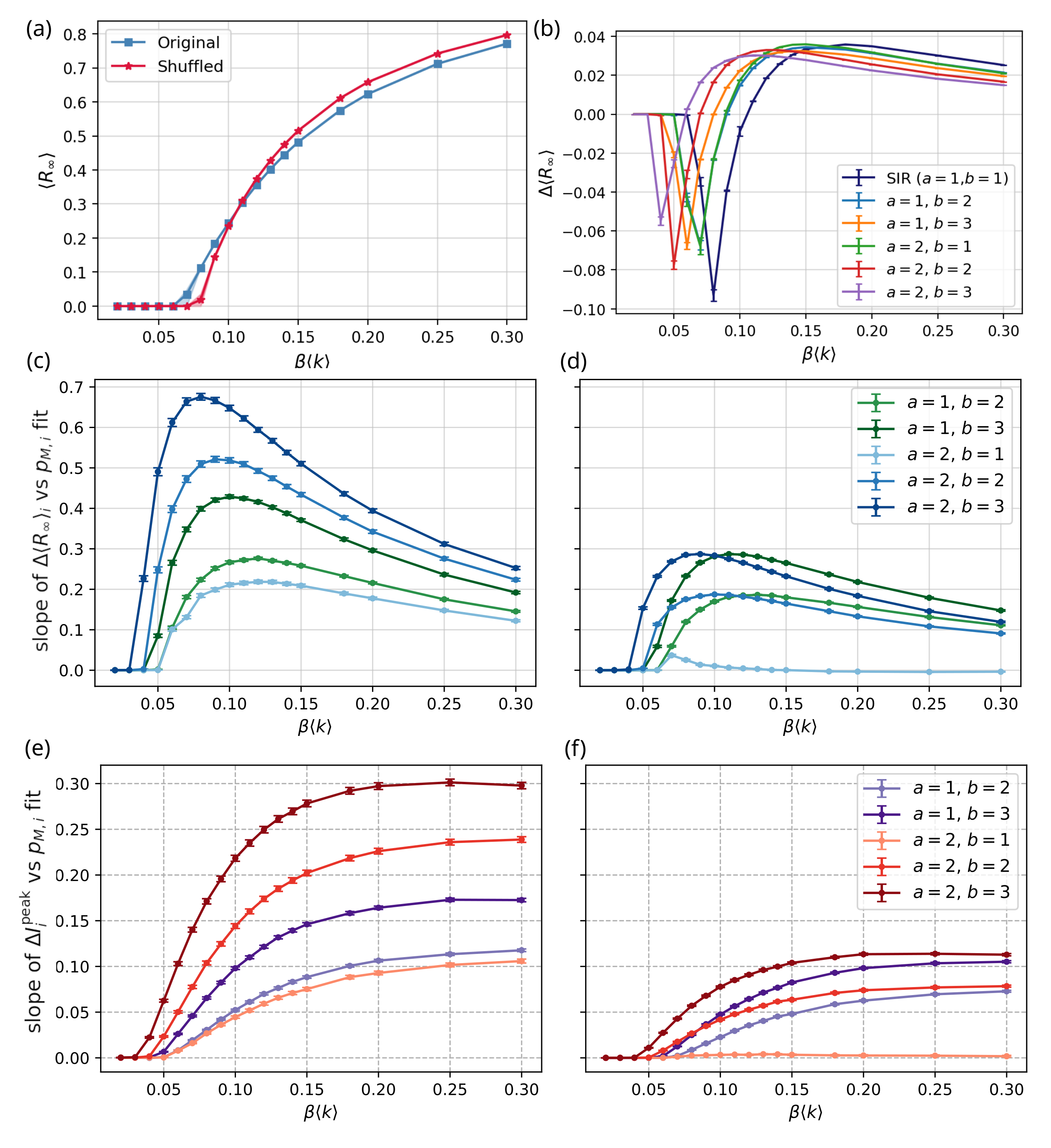}
\caption{
	Comparison of results using original and shuffled contact graphs in the city of Milan. 
	Panel (a): Attack Rate (AR) of the SIR model for both graphs. 
	Panel (b): Difference in AR (shuffled graph minus original graph) for the HeSIR and SIR models across various values of \(a, b\). 
	Panels (c-d): Comparison of the slopes of the linear fit of \(\Delta \left< R_{\infty} \right>_i\) versus \(p_{M,i}\) as a function of \(\beta\), for different \(a, b\) values in the HeSIR model with data-driven distribution. On the left, the slopes for the original graph, on the right the slopes for the shuffled graph.
	Panels (e-f): Comparison of the slopes of the linear fit of $\Delta I^\textrm{peak}_{i}$ versus $p_{M,i}$, for different \(a, b\) values in the HeSIR model with data-driven distribution, analogously as panel (c).
}
\label{fig:confmod_results}
\end{figure}

We replicate the analyses from the previous sections using the shuffled contact graph, fixing the fraction of non-compliant individuals at \( p_M = 0.2 \). 
Under a uniform distribution of M individuals, we observe that the epidemic threshold \(\beta\) for an outbreak is higher in the shuffled graph, but outbreaks become more severe at higher \(\beta\) values, as illustrated in Figure \ref{fig:confmod_results} (top row). 
In other words, fewer infections occur at low \(\beta\) with the shuffled graph, whereas beyond a certain \(\beta\) the number of infections surpasses that of the original graph. This pattern holds for both the baseline SIR model (\(a=b=1\)) and the HeSIR model, although the effect is less pronounced in the latter. 
This is evident in Figure \ref{fig:confmod_results} (top-right panel), which displays the increase in attack rate when using the shuffled contact graph relative to the original. Overall, the observed differences remain modest.

The center and bottom rows of Figure \ref{fig:confmod_results} compare the results obtained with the HeSIR model using the data-driven and uniform distributions of non-compliant individuals, as discussed in the previous section. 
Here, however, we analyze the impact of using the shuffled contact graph (left column) instead of the original graph (right column).
We observe a significant reduction in the dependence of the excess AR and the infection peak on \( p_{M,i} \), the fraction of misbehaving individuals in each tile: the slope of the linear fit decreases by at least a factor of two for the AR and by a factor of three for the peak amplitude. 
More specifically, increasing the infectivity parameter \( a \) appears less influential than increasing susceptibility \( b \), as the curves change little when varying \( a \) compared to \( b \). 
Notably, when \( b = 1 \), both curves are close to zero, indicating that the dependence of the excess AR and infection peak on \( p_{M,i} \) is almost eliminated. 
This effect arises from the edge shuffling process, which causes most connections of an individual within a tile to link with individuals in other tiles. 
In contrast, in the original graph, about 20\% of edges connect individuals within the same tile. 
Consequently, geographical areas with a high concentration of misbehaving individuals lose their role as hotspots of local transmission, and the disruption of the network structure reduces spatial heterogeneity in the epidemic spread.

\section{Discussion}

In this study, we investigated the impact of non-compliant individuals on epidemic propagation within urban environments. 
Leveraging detailed geographic data, we constructed large-scale contact networks---comprising up to 1.2 million nodes---for three Italian cities, explicitly distinguishing between household-based and broader social interactions. 
Our minimally modified SIR framework incorporates a subpopulation of non-compliant individuals, which leads to a substantial increase in the total number of infections and a marked acceleration of epidemic dynamics. 
The resulting amplification of the infection force enables outbreaks to occur even for diseases with lower intrinsic transmission rates.
Moreover, we find that the peak number of infections rises significantly when non-compliant behaviour is more pronounced, even if such individuals represent only a small fraction of the urban population. 
These results underscore the critical role that behavioural heterogeneity—specifically, non-compliance—can play in facilitating the spread of infectious diseases.

Building on empirical links between health-related non-compliance and political affiliation, we constructed a data-driven, geographically heterogeneous distribution of non-compliant individuals and compared its epidemiological effects to those of a homogeneous distribution. 
While city-level epidemic metrics, such as attack rate, peak time, and peak amplitude, showed limited but non-negligible differences between the two scenarios, substantial disparities emerged at finer spatial scales, i.e., at the tile level.

We observed that both the local attack rate and epidemic peak were strongly influenced by the local concentration of non-compliant individuals. 
As expected, when the spatial distribution of non-compliant individuals is heterogeneous, the epidemic tends to intensify in regions with a higher density of non-compliant individuals and remains comparatively subdued in areas with lower density. 
Notably, this spatial dependence becomes more pronounced when the disease's baseline transmission rate is just above the epidemic threshold, indicating that behavioural heterogeneity exerts a greater influence on marginal outbreak conditions.

When comparing these results to those obtained using a shuffled version of the contact graph, we observed only marginal differences at the city-wide scale, but substantial effects at the local (tile) level. 
In particular, the observed dependence of local attack rates and epidemic peaks on the fraction of non-compliant individuals was significantly reduced for certain values of the misbehaviour parameters, and in some cases nearly eliminated.
These findings highlight the critical role played by the two-level city contact graphs employed in our study, which maintain a realistic balance between local (household) and long-range (social) connections. 
Randomly shuffling edges disrupts this structure, altering the ratio of internal to total connections within tiles and thereby modifying the network's spatial organization. 
This disruption increases the number of potential transmission pathways, reducing the influence of localized behavioural clustering and diluting the spatial heterogeneity that would otherwise shape epidemic outcomes.

In summary, our results demonstrate that the presence of non-compliant individuals can significantly alter epidemic dynamics, even when they constitute only a small fraction of the overall population. 
Their spatial concentration within specific urban areas is particularly concerning, as it can generate infection ``hotspots'' that place disproportionate pressure on local healthcare infrastructure. 
These findings highlight the potential value of geographically targeted public health interventions, both for epidemic preparedness and for behaviour-focused campaigns aimed at increasing adherence to public health guidelines, including but not limited to mask usage, physical distancing, and vaccination.

Comparisons between the data-driven and uniformly random spatial distributions of non-compliant individuals revealed only modest differences at the city-wide level, though these were measurable. 
This suggests the presence of additional mechanisms, potentially related to geography or network structure, that merit further investigation. 
In particular, it remains an open question to what extent the location of clusters of non-compliant individuals influences macro-scale epidemic outcomes.

While our work relies on generating of large contact graphs and describing the state of each agent/individual, a well-known alternative approach to studying urban epidemics is based on metapopulation models, in which individuals are distributed across different areas (``patches``) within the city and can move from one patch to another during the simulation ~\cite{hazarie_interplay_2021}. This approach may involve a lighter computational load, but it relies on the assumption that the population of a patch is well-mixed, meaning that any individual can be infected by any other. More importantly, metapopulation models are better suited to including mobility in the model: however, we did not have access to such data for this study. Ultimately, we chose the modelling approach put forward by~\cite{guarino_inferring_2021} as it provides a proven method for building urban social networks from publicly available datasets. 

Our method for estimating the spatial distribution of non-compliance was based on political affiliation data and vaccine hesitancy adaptation coefficients. 
While this provides a plausible proxy for health-related behavioural resistance, it captures only a subset of the broader phenomenon. 
Future refinements could incorporate socio-economic and demographic variables to improve the realism of misbehaviour propensity models.
Additionally, our simulations assumed random initial infections, with no preferential spatial placement of index cases. 
Alternative scenarios, in which outbreaks originate in areas with high numbers of non-compliant individuals or at geographically critical nodes such as transportation hubs, could yield different dynamics and will be explored in future work. 
Finally, our model considers behaviour as static, whereas real-world individuals often adapt their actions during an epidemic, for example by reducing contacts in response to perceived risk \cite{mao_modeling_2014}. 
Incorporating such dynamic behavioural feedback mechanisms remains an important direction for future research.


\paragraph{Data Access}{The Julia code used for the simulations is available at~\cite{JuliaCode}, while the non-compliance distributions obtained for the cities are available in the Github repo~\cite{noncompliance_data}.}

\paragraph{Acknowledgments}{This work was supported by the project “CODE – Coupling Opinion Dynamics with Epidemics”, funded under PNRR Mission 4 "Education and Research" - Component C2 - Investment 1.1 - Next Generation EU "Fund for National Research Program and Projects of Significant National Interest" PRIN 2022 PNRR, grant code P2022AKRZ9, CUP B53D23026080001.}

\section*{Appendix A. Construction of the contact graph} 

The city contact graph $\mathcal{G}(V,E)$ is actually composed of two subgraphs that share the same set of nodes $V$, corresponding to the set of individuals. The subgraph $\mathcal{G}_H (V,E_H)$ defines the set of household contacts, while $\mathcal{G}_F (V, E_F)$ defines the set of social contacts, and $E=E_H \cup E_F$.

\begin{figure}[t]
\centering
\includegraphics[height=0.3\linewidth]{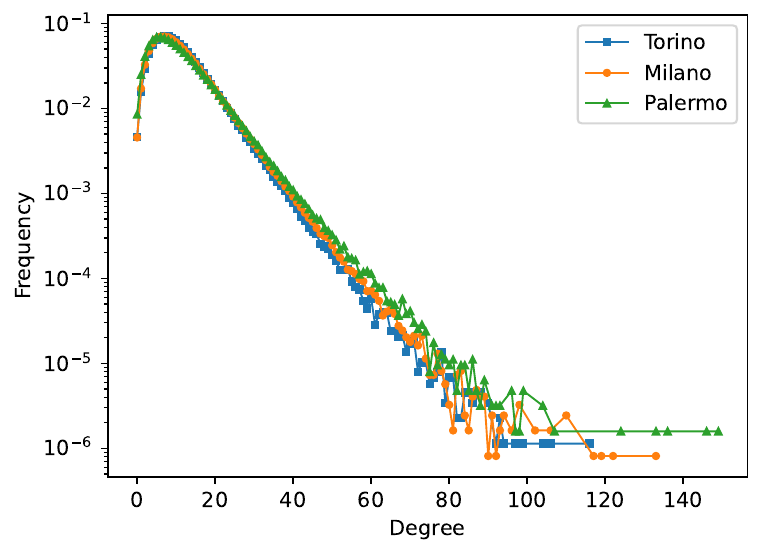}
\includegraphics[height=0.3\linewidth]{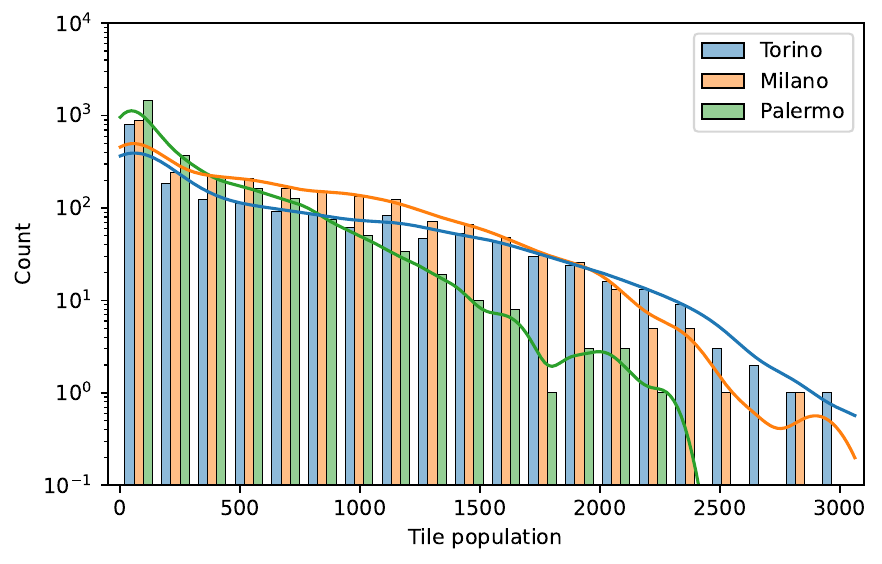}
\caption{Left: degree distribution of the contact graph generated for the cities of Torino, Milano, and Palermo. Right: distribution of the number of individuals per tile (the continuous lines are the density estimates and serve as a visual guide only).}
\label{fig:graphs_stats}
\end{figure}

As introduced in Sec. \ref{sec:graph_model}, the city area is divided into $N_T$ square tiles of fixed size, within which individuals are positioned according to the WorldPop distribution~\cite{worldpop}. Each individual belongs to one of the following age groups: children (under 18 years old), young adults (18 to 34), adults (35 to 64) or elderly (65 and above). 
The social contacts graph $\mathcal{G}_F$ takes into account friendships and frequent contacts and is built starting from the Fitness-Corrected Block Model (FCBM)~\cite{bernaschi_fitness-corrected_2022}, with one block for each tile and age group. To quantify interactions between age groups, a matrix $S$ is defined, whose entry $s_{gh}$ measures the frequency of social contacts between age groups $g$ and $h$.
Furthermore, the probability of contact between two individuals $u$ and $v$ belonging, respectively, to tiles $i$ and $j$ should be proportional to $d_{ij}^{-2}$, where $d_{ij}$ is the geographical distance between the centers of the respective tiles, divided by half of the tile side ($d_{ii}=1$ for nodes in the same tile). With the definition of fitness $f_u$ for every individual $u$, drawn from a shifted lognormal distribution $f_u \sim 1+ \mathrm{Lognormal}\left(\ln (2), 0.25 \right)$, the probability of a contact between individuals $u$ and $v$ can be written as:
\begin{equation}\label{eq:uvprob}
\mathrm{Prob}\left[u,v\right]=\frac{\mu N}{2} \frac{m_{IJ}\ s_{IJ}}{\sum_{i\leq j}\left(m_{ij} s_{ij}\right)}\frac{d_{uv}^{-\gamma}f_{u}f_{v}}{\sum_{u'\in V_{I},v'\in V_{J}}\left(d_{u'v'}^{-\gamma} f_{u'}f_{v'}\right)}
\end{equation}
where $\mu$ is the required average number of contacts per node, $N$ is the number of nodes, $I$ and $J$ are the indices of the age blocks of, respectively, $u$ and $v$, $m_{IJ}$ the number of possible pairs of nodes in block $I$ and $J$, and $d_{uv}=d_{ij}$ is the aforementioned distance between the respective tiles of $u$ and $v$. In Figure~\ref{fig:graphs_stats} we report the degree distribution of the contact graphs generated for the three cities studied in this work, as well as the distribution of the number of individuals per tile.

\vskip2pc

\bibliographystyle{RS}
\bibliography{literature}

\end{document}



\title{Supplementary Material \\ \textit{A data-driven analysis of the impact of non-compliant individuals on epidemic diffusion in urban context}}

\author{
Fabio Mazza, Marco Brambilla, Carlo Piccardi, and Francesco Pierri}
\affil{Dipartimento di Elettronica, Informazione e Bioingegneria, Politecnico di Milano, Piazza Leonardo da Vinci 32, 20133 Milano, Italy}

\date{} 

\maketitle

\section{Additional statistics on the contact graphs}
\begin{figure}[h]
	\centering
	\includegraphics[width=0.5\linewidth]{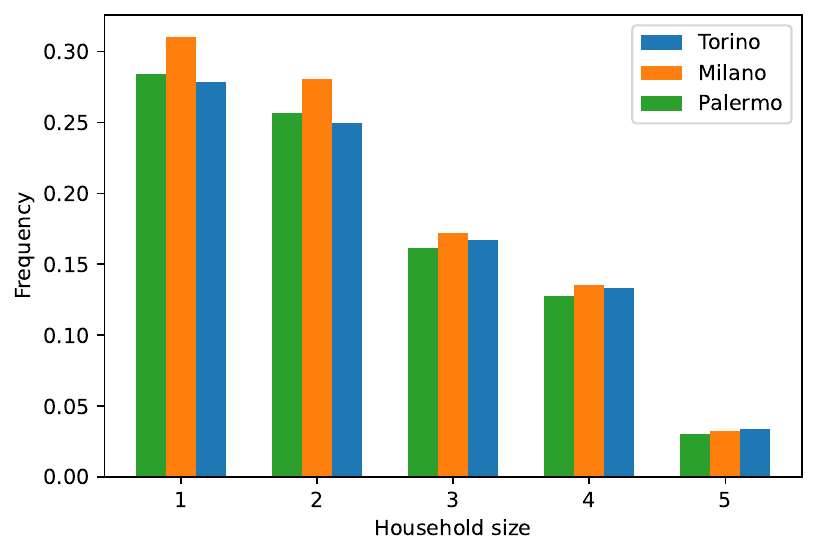}
	\caption{Distribution of the household size for each city contact network generated}
	\label{fig:household}
\end{figure}

Figure \ref{fig:household} shows the distribution of household sizes in the contact graphs generated, reflecting Italian census data from ISTAT\footnote{ISTAT, Italian National Institute of Statistics, \url{https://www.istat.it/en/}}.

\section{Results with uniform distribution of non-compliant individuals}
\begin{figure}[h]
	\begin{centering}
    \includegraphics[width=0.85\columnwidth]{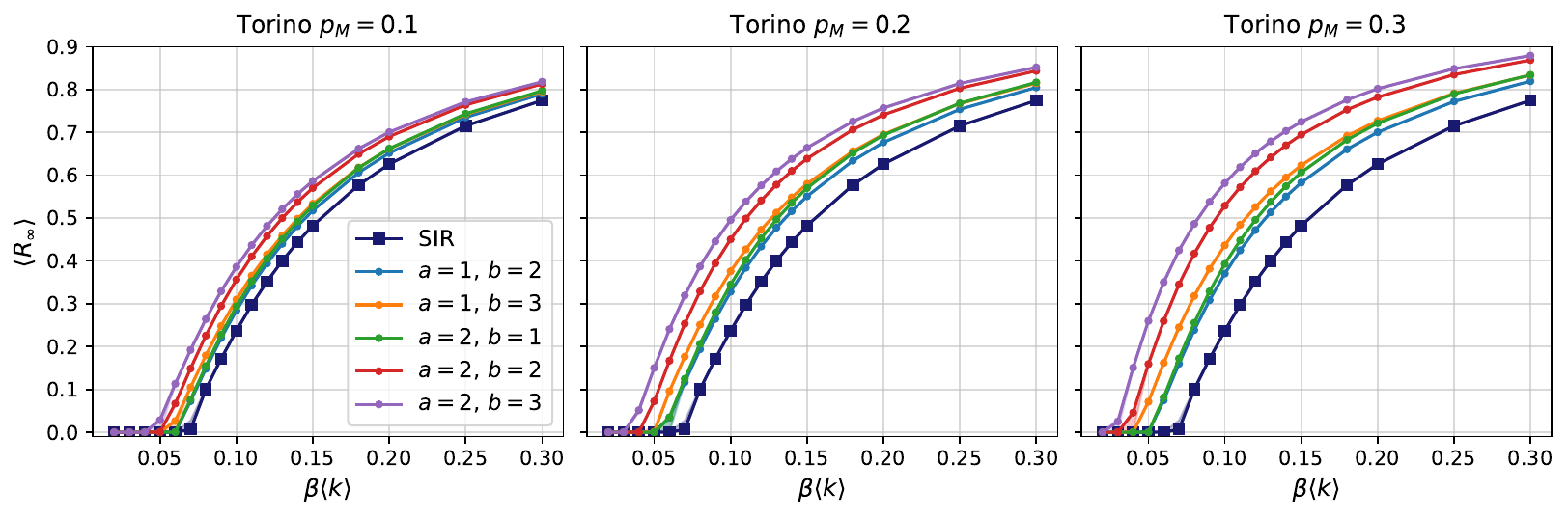}
		\par\end{centering}
	
	\begin{centering}
	\includegraphics[width=0.85\columnwidth]{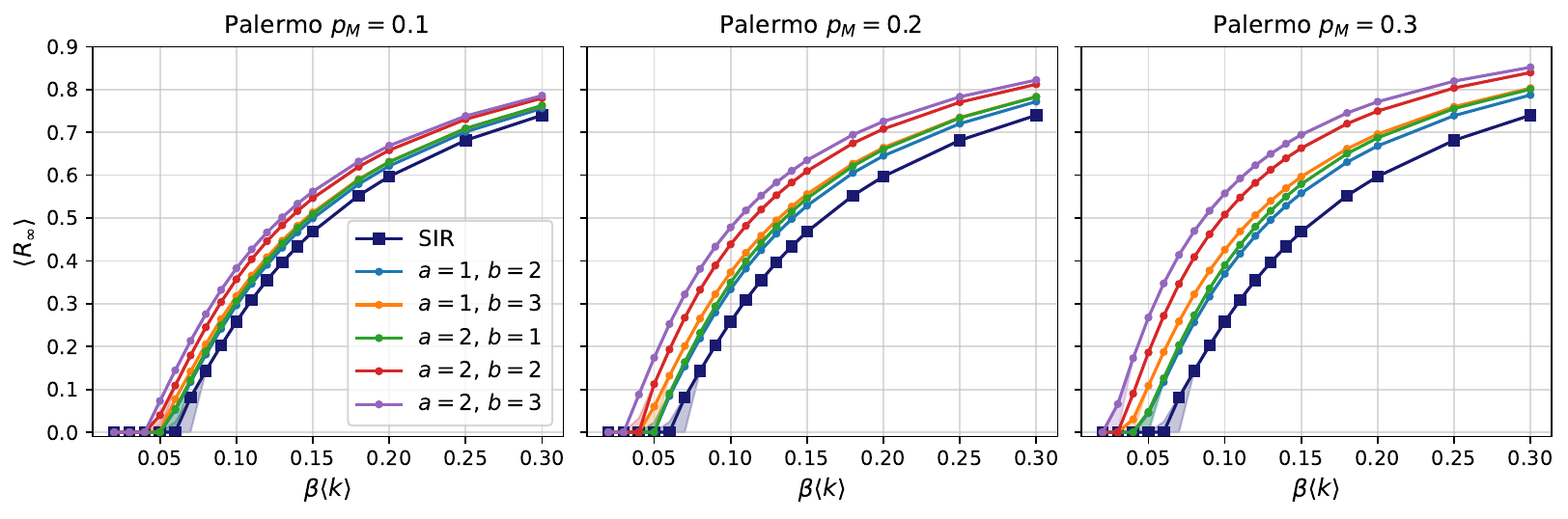}
		\par\end{centering}
	\caption{ Simulations of the HeSIR model in the cities of Torino and Palermo with a uniform distribution of non-compliant individuals: the panels report the total fraction of infected individuals (Attack Rate) as a function of the product of transmission rate $\beta$ and average graph degree $\langle k \rangle$, for varying values of $a$, $b$, and $p_M$. In all panels, lines indicate the mean and shaded areas
correspond to inter-quartile range. }\label{fig:AR_cities_extra}
\end{figure}

\begin{figure}[ht]
	\begin{centering}
\includegraphics[width=0.85\columnwidth]{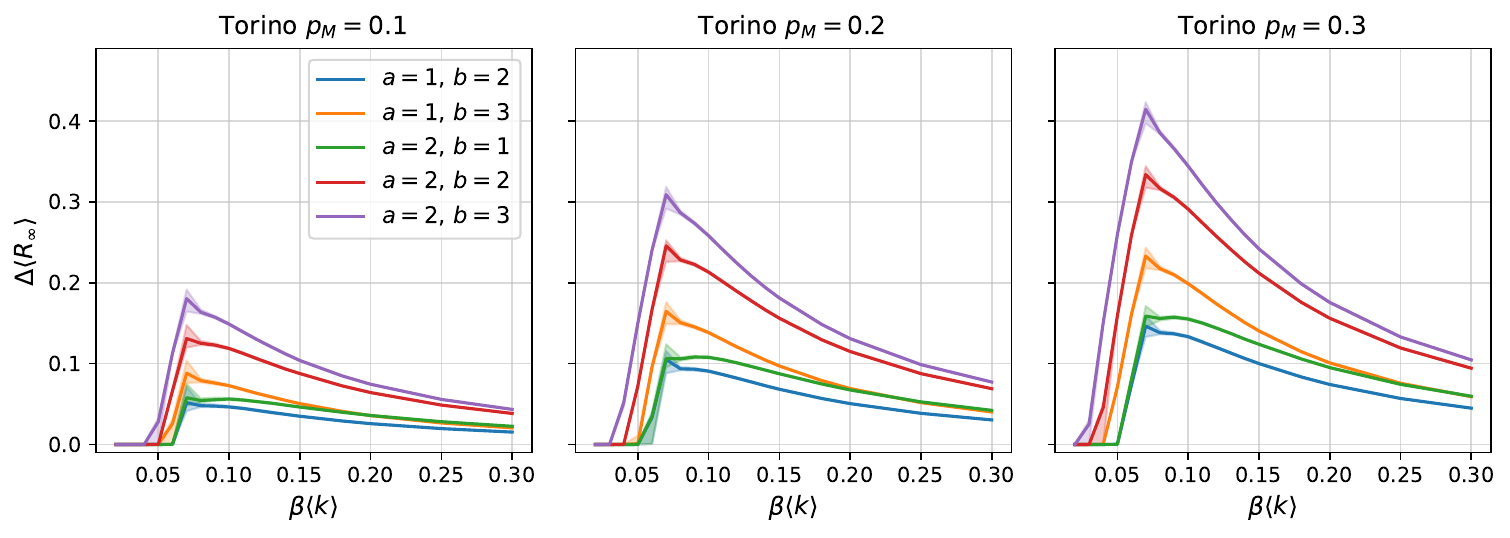}
		\par\end{centering}
	\begin{centering}
	\includegraphics[width=0.85\columnwidth]{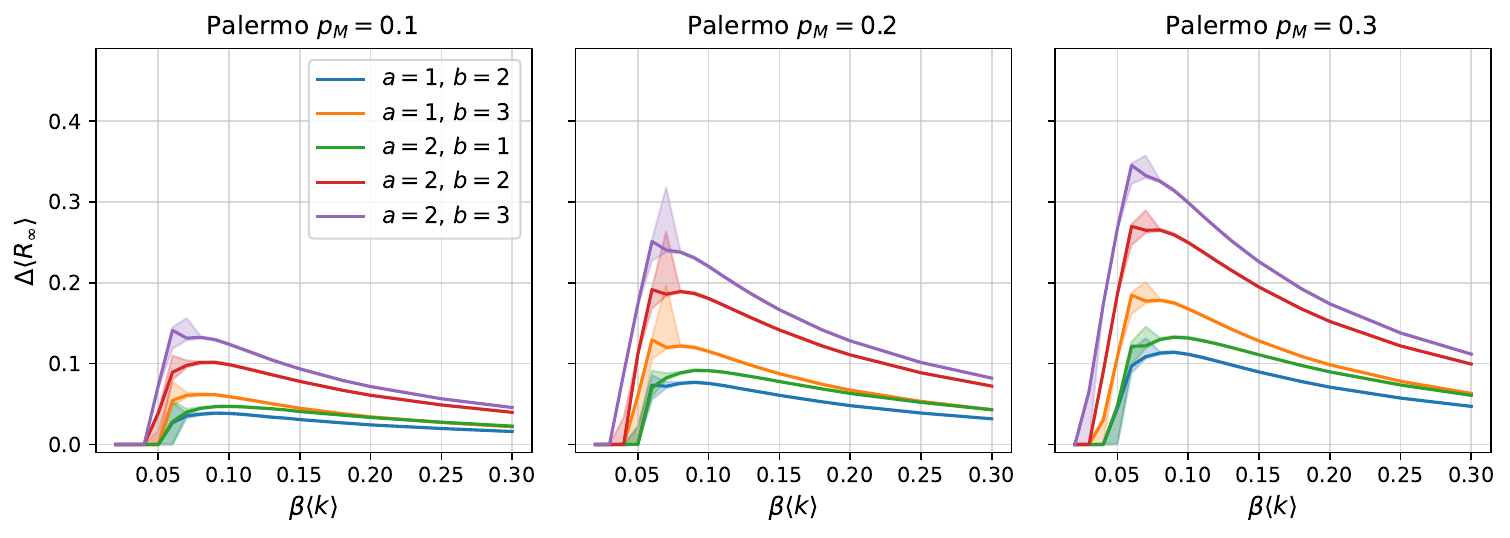}
		\par\end{centering}
	\centering{}\caption{Simulations of the HeSIR model in the cities of Torino and Palermo with a uniform distribution of non-compliant individuals: the panels report the difference in Attack Rates between the HeSIR and baseline SIR models. In all panels, lines indicate the mean and shaded areas
correspond to inter-quartile range. }
	\label{fig:AR_diff_extra}
\end{figure}

Figures \ref{fig:AR_cities_extra} and \ref{fig:AR_diff_extra} show the attack rate curves and the difference in attack rates between the HeSIR
and SIR models, respectively. As can be seen from these plots, the qualitative behaviour is the same
across all three cities considered.

As can be seen in Figure \ref{fig:AR_cities_extra}, Palermo generally shows slightly lower attack rates (approximately 3\%), in both the SIR and HeSIR models.
However, we can see from Figure \ref{fig:AR_diff_extra} that the difference in city Attack Rate between the HeSIR and SIR model is higher in Torino than in Palermo, especially for large quantities of M individuals ($p_M=0.3$), for which there are around 5\% more infected.


\clearpage

\section{Results with data-driven distribution of non-compliant individuals}

\begin{figure}[t]
\centering
    \includegraphics[width=0.84\columnwidth]{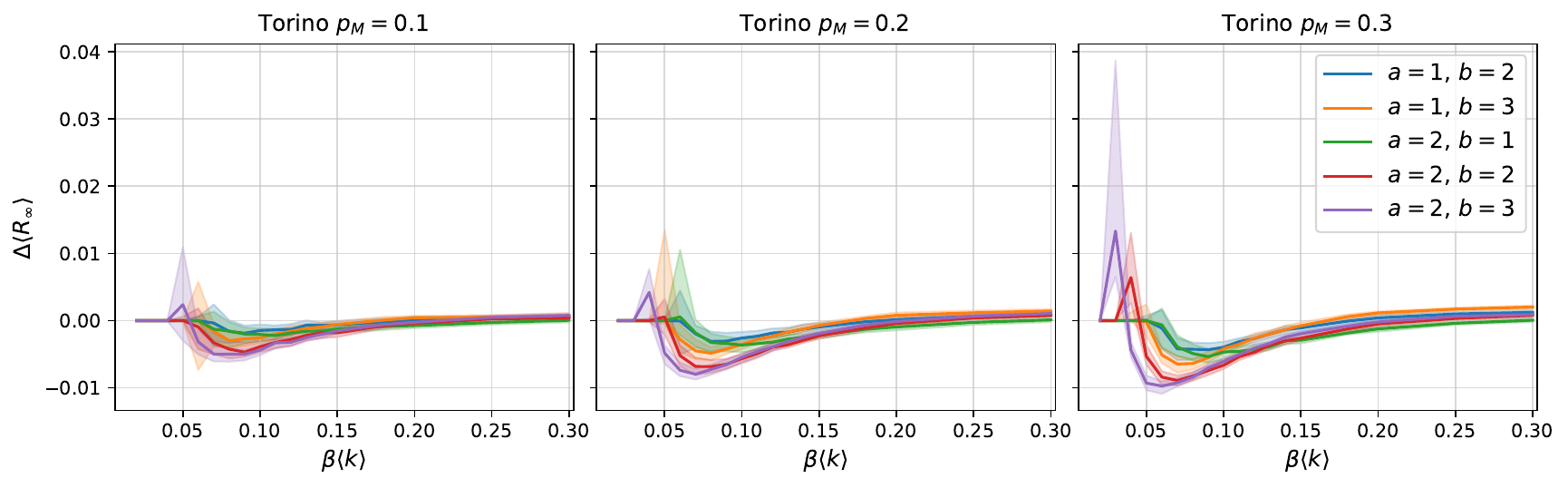}\\
    \includegraphics[width=0.84\columnwidth]{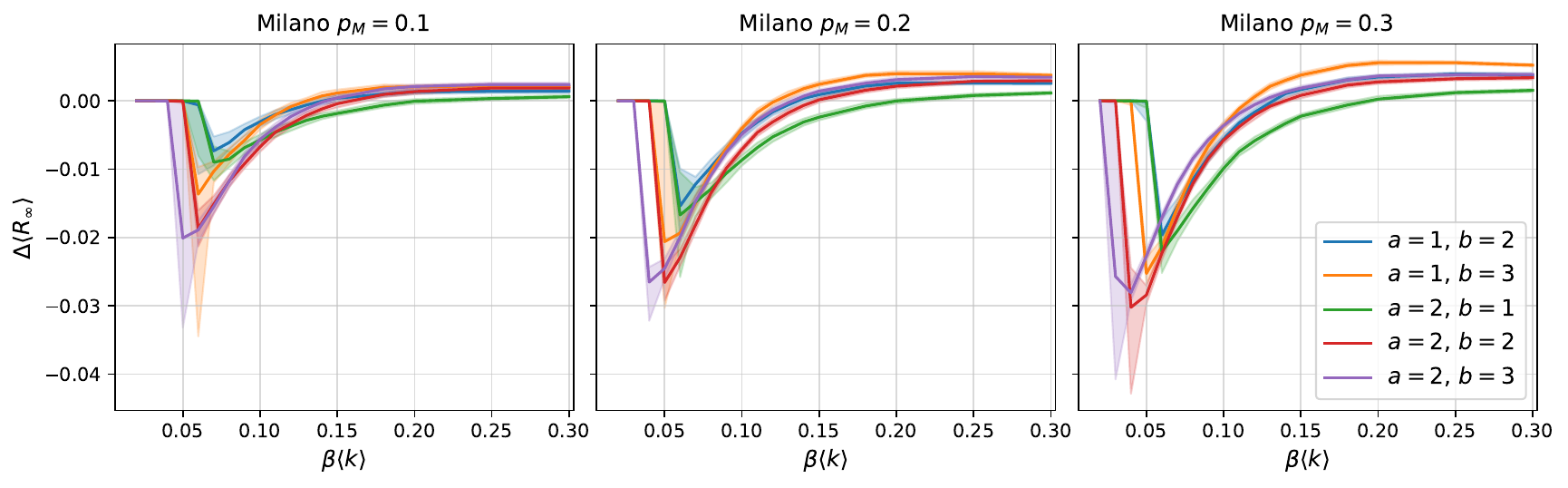}\\
    \includegraphics[width=0.84\columnwidth]{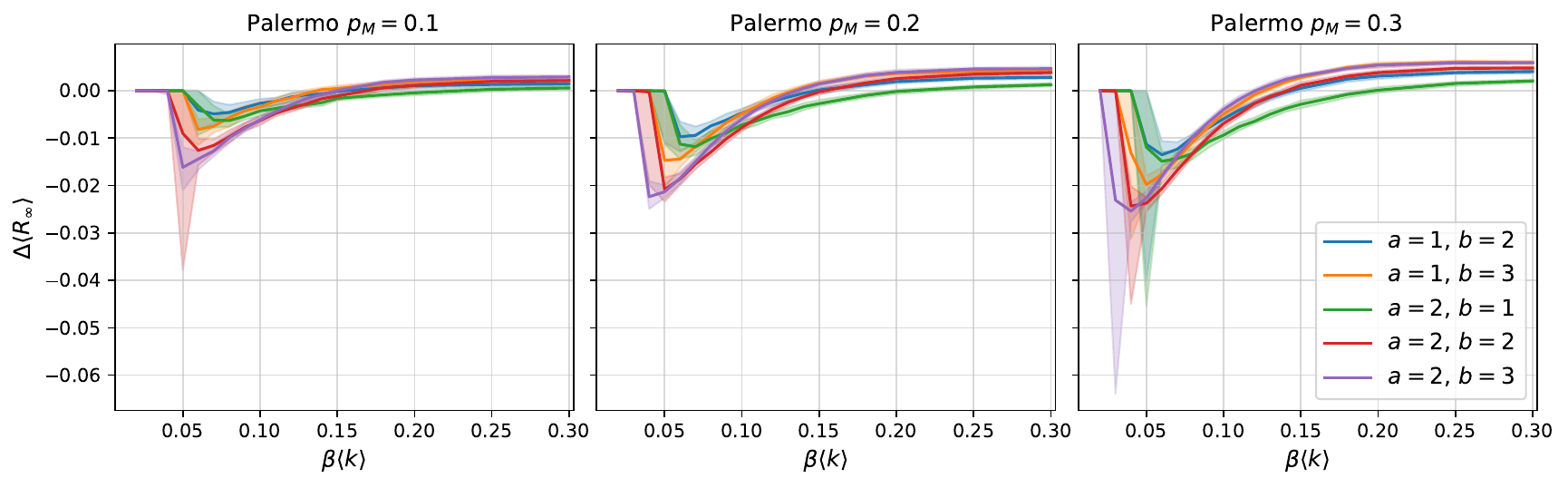}\\
		\caption{  Difference in Attack Rate in the cities of Torino, Milano, and Palermo between the data-driven and uniform distributions of non-compliant
individuals, for different fraction of non-compliant individuals $p_M$. Negative values indicate fewer infections under the
data-driven distribution, and positive values indicate more.
Lines indicate the median value, and shaded areas the interquartile range from the simulations.
}\label{fig:ARdiff}
\end{figure}

In this section, we analyse the differences in attack rates across all the cities considered (Figure \ref{fig:ARdiff}). While these differences are small, reaching no more than 5\% of the total population for specific values of $\beta$, there is a measurable qualitative difference between the cities.
For low to intermediate values of $\beta$, the AR difference is negative, meaning fewer people are infected for the data-driven distribution in Milano and Palermo, while the opposite is true for Torino (although the difference is smaller in absolute value).

\begin{figure}[t]
	\begin{centering}
    \includegraphics[width=0.46\columnwidth]{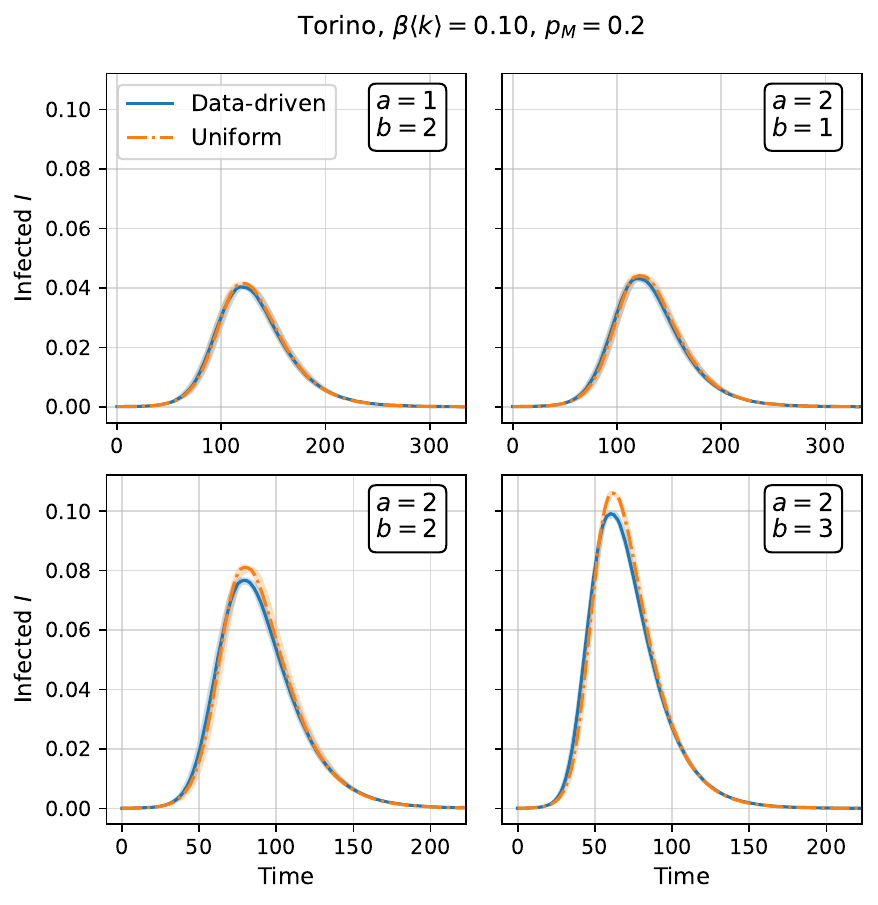}
		\includegraphics[width=0.46\columnwidth]{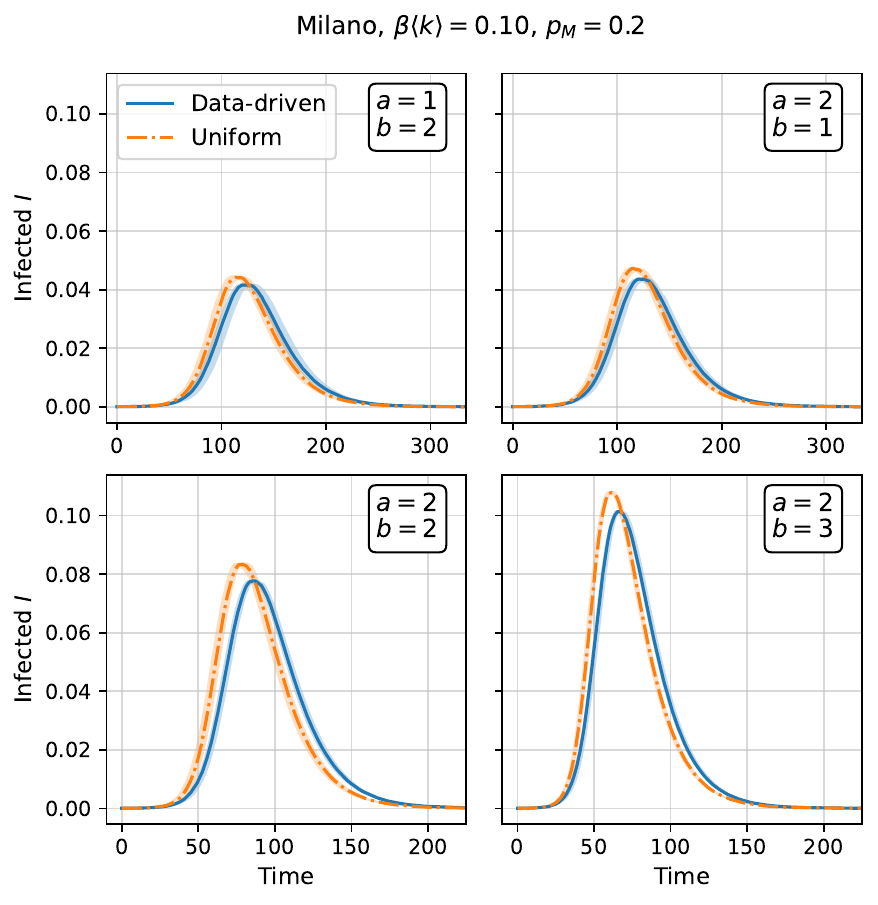}
		\par\end{centering}
	\caption{   
    Epidemic outbreak simulations of the HeSIR model in the cities of Torino and Milano with uniform and data-driven distribution of non-compliant individuals.
    Lines indicate the median value, and shaded areas the interquartile range from the simulations.}\label{fig:Iplot_compare}
\end{figure}

We also observe a small but significant difference when comparing the dynamics across different cities. Figure \ref{fig:Iplot_compare} shows the number of infected $I$ during an oubreak in Torino and Milano, for different values of $a$,$b$ and fixed $\beta$ and $p_M$. We can observe that for both cities the peak of the infected is slightly higher for the uniform distribution. However, in Milano the peak occurs earlier in the uniform case than in the data-driven case, while in Torino the peak times are similar. 

\begin{figure}[t]

\begin{centering}
	\includegraphics[width=0.32\columnwidth]{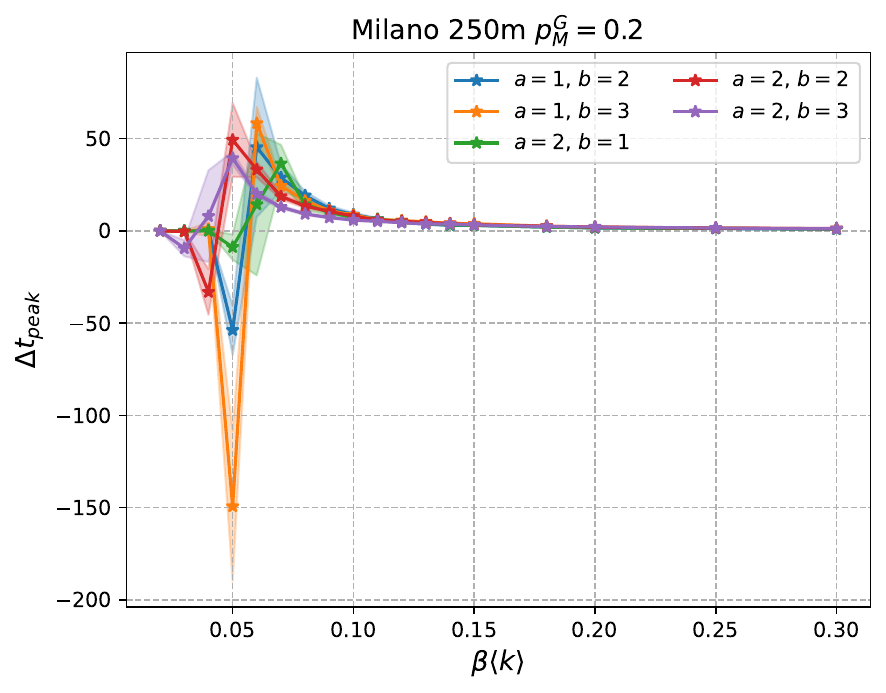}\includegraphics[width=0.32\columnwidth]{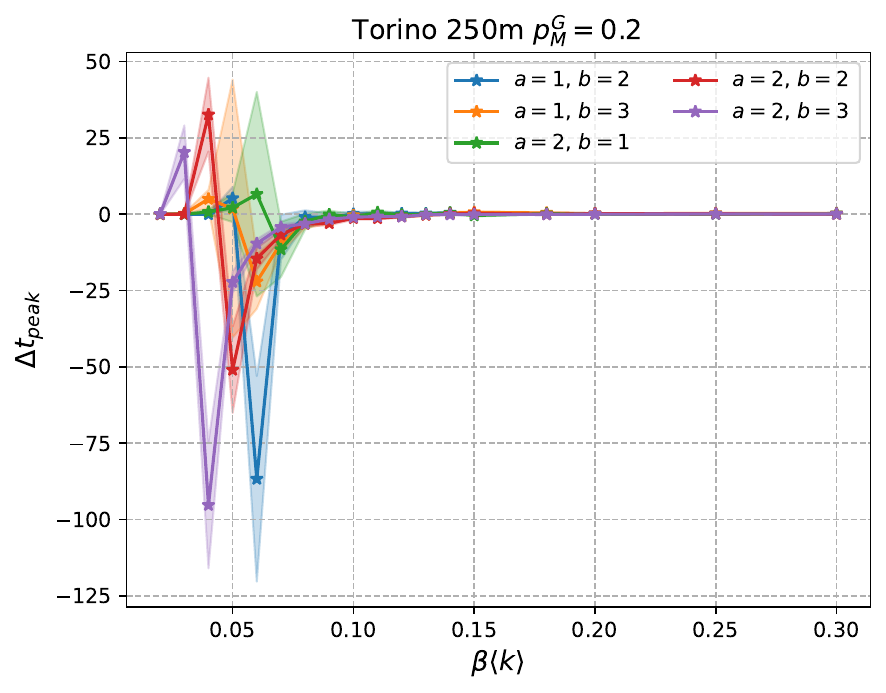}
	\par\end{centering}
\caption{
Difference in peak times of the infecteds between data-driven and uniform distribution, for different values of $\beta$, $a$, $b$. A positive difference means that the epidemic peaks arrives later in the data-driven case.}\label{fig:peaktimes_diff_all}
\end{figure}

Regarding the peak time of the epidemic, the difference between the
data-driven and the homogeneous case is vanishingly small for $\beta\left\langle k\right\rangle >0.1$.
As shown in Figure \ref{fig:peaktimes_diff_all}, for lower values of the infectivity and above the epidemic threshold,
in Milano the difference in peak time $\Delta t_{\text{peak}}$ becomes positive after a brief
transition, meaning that the peak arrives later in the data-driven
case. In Torino, instead, $\Delta t_{\text{peak}}$ has the opposite behaviour,
indicating that the peak arrives sooner in
the data-driven case.

\clearpage

\section{Differences at tile-level between uniform and data-driven distribution of non-compliant individuals}

\begin{figure}
	\begin{centering}
		\includegraphics[width=0.75\columnwidth]{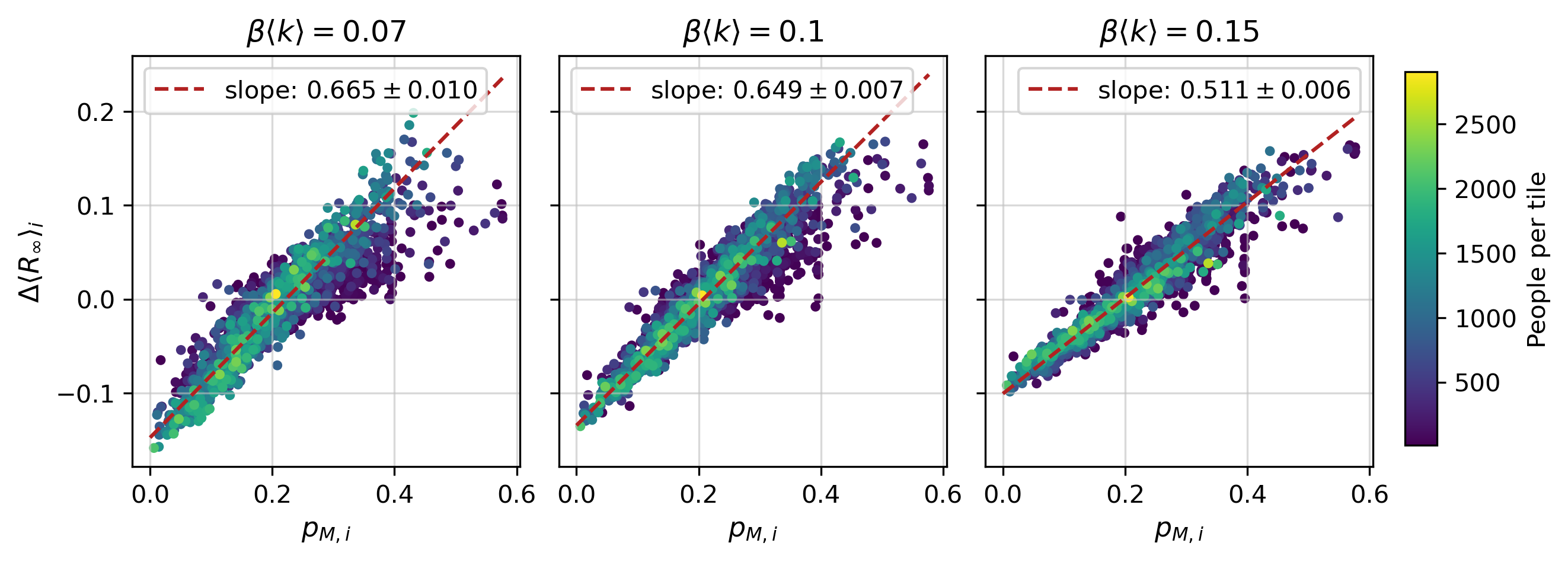}
		\par\end{centering}
	\begin{centering}
		\includegraphics[width=0.75\columnwidth]{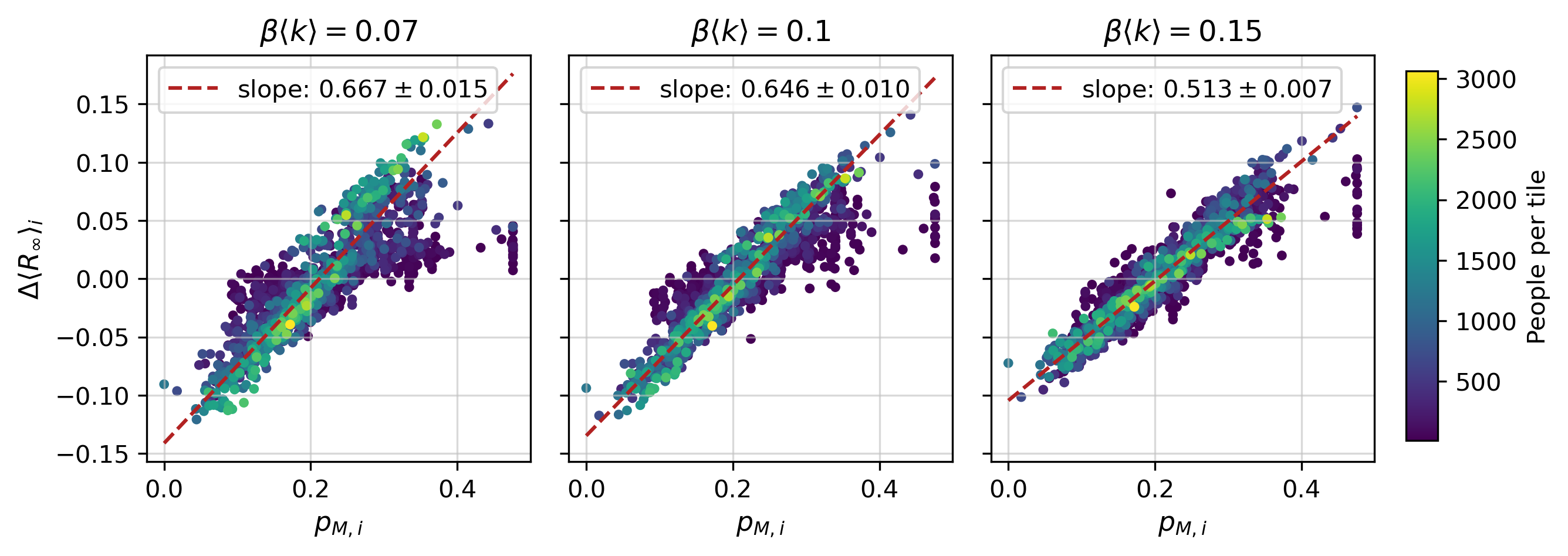}
		\par\end{centering}
	\begin{centering}
		\includegraphics[width=0.75\columnwidth]{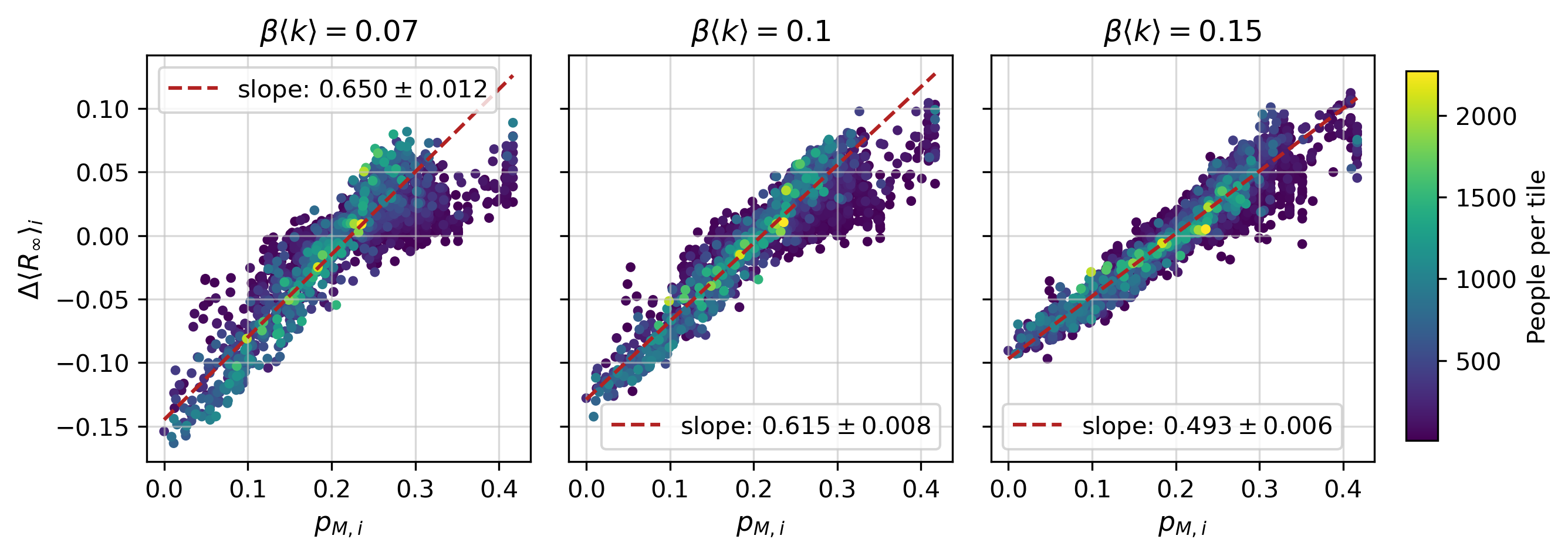}
		\par\end{centering}
	\caption{   
    Simulations with the HeSIR model in the cities of Torino (top row), Milano (middle row), and Palermo (bottom row). The scatter plots show the dependence of $\Delta \left< R_{\infty}\right>_i$ on the fraction $p_{M,i}$ of non-compliant individuals (the city average is $p_M=0.2$). Each dot corresponds to a tile, whose population is color-coded (see colorbar on the right). The red dashed line is the best linear fit (weighted by the tile population): the slope value (with 95\% CI) is reported at the top of the panel. The other simulation parameters are set to $a=2$, $b=3$.}\label{fig:DeltaAR_tiles_all}
\end{figure}

\begin{figure}[h]
	\begin{centering}
    \includegraphics[width=0.325\columnwidth]{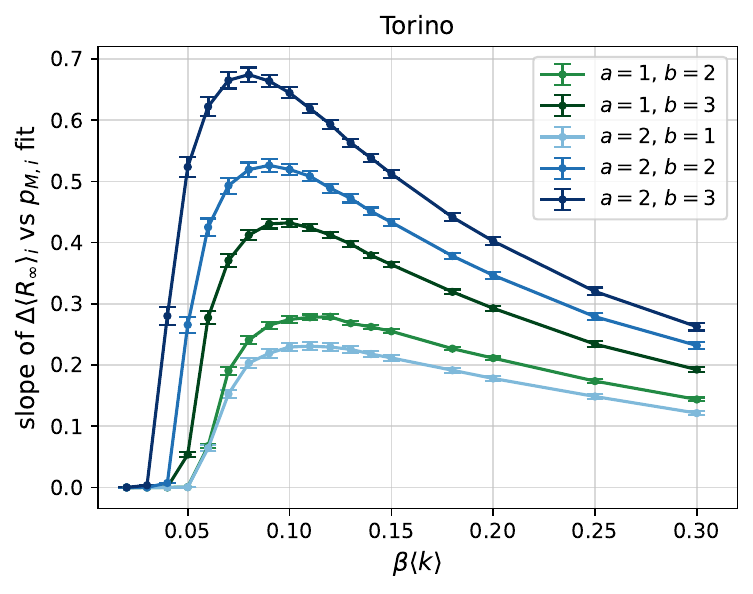}
    \includegraphics[width=0.325\columnwidth]{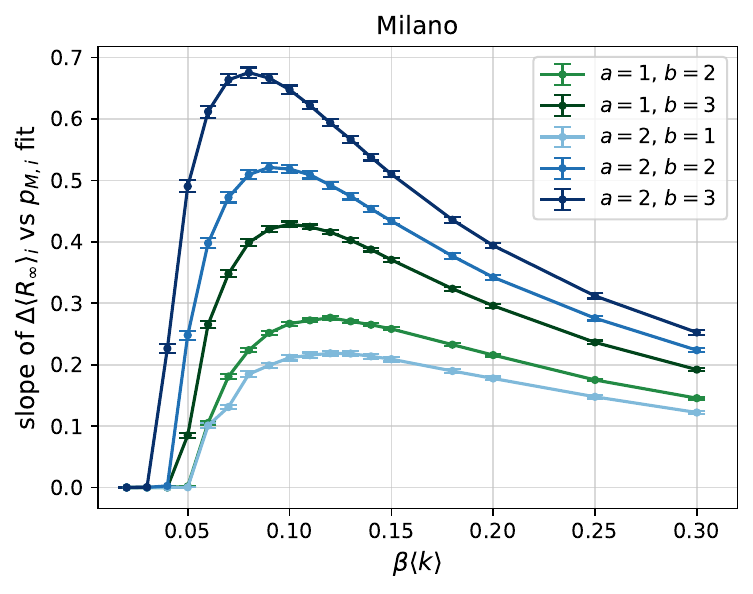}
        \includegraphics[width=0.325\columnwidth]{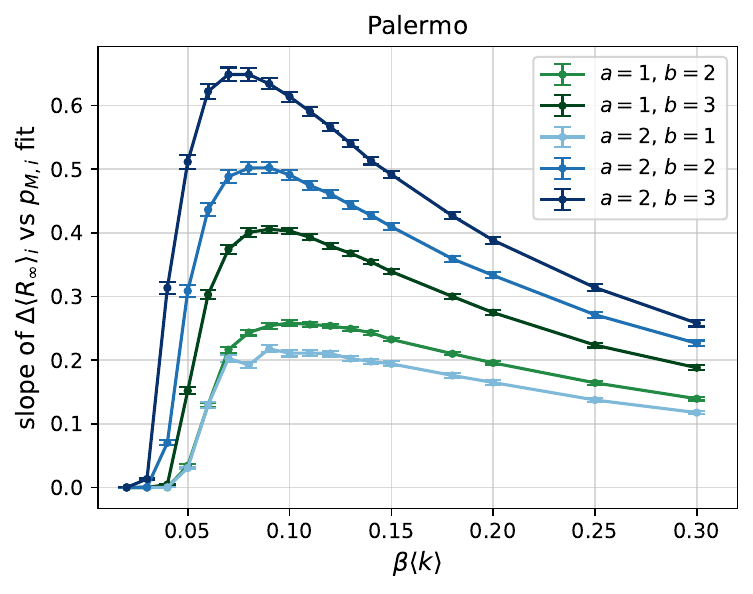}
		\par\end{centering}
	\centering{}\caption{ Simulations with the HeSIR model in the cities of Torino, Milano, and Palermo, using the data-driven distribution. The panels report the slope of the best linear fit of $\Delta \left< R_{\infty}\right>_i$ versus $p_{M,i}$ as a function of $\beta$, for various values of $a$ and $b$ and fixed $p_M=0.2$ (see Figure \ref{fig:DeltaAR_tiles_all}).  Each point corresponds to a simulation set used for fitting; error bars represent the 95\% confidence intervals.  }\label{fig:slopes_AR_tiles_all}
\end{figure}

\begin{figure}[t]
    \begin{centering}
		\includegraphics[width=0.326\linewidth]{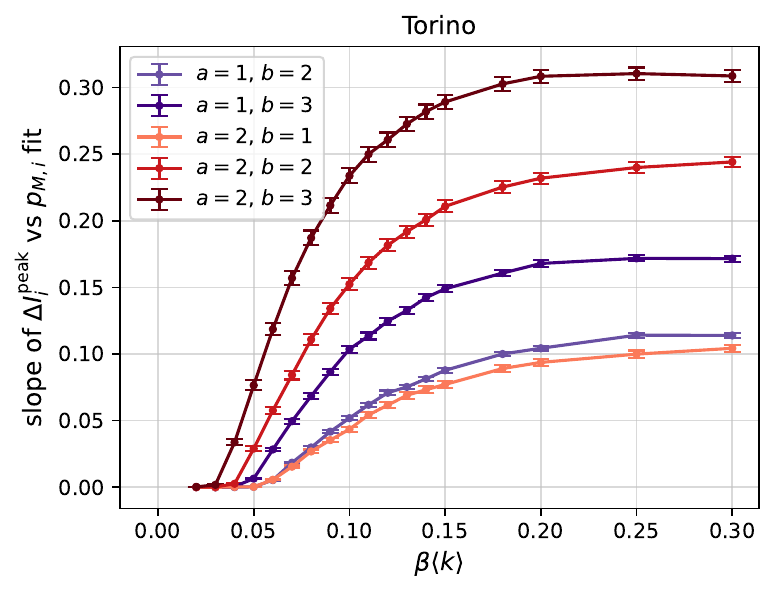}
		\includegraphics[width=0.326\linewidth]{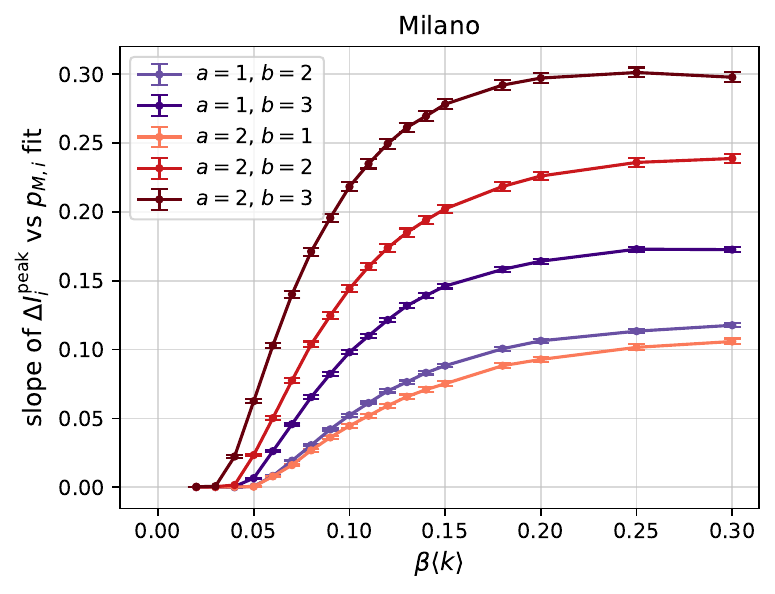}
		\includegraphics[width=0.326\linewidth]{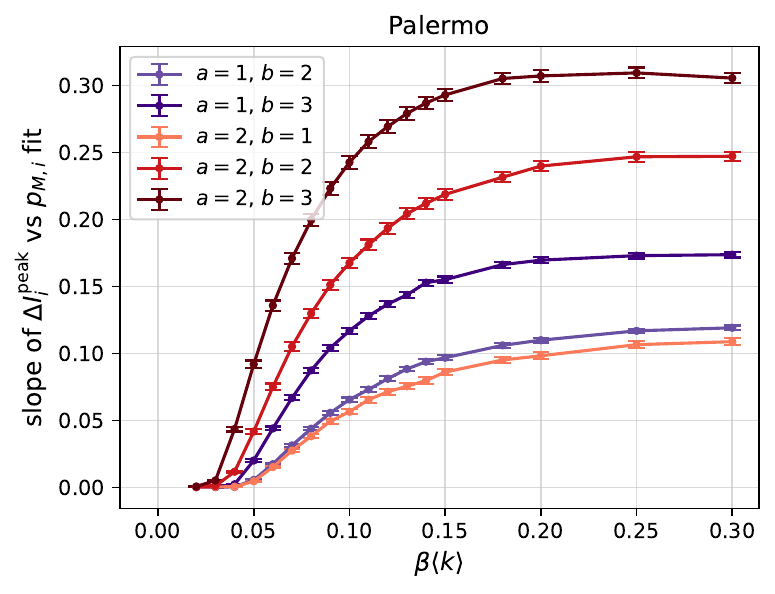}
		\par\end{centering}
	\caption{   
   Simulations with the HeSIR model in the cities of Torino, Milano, and Palermo, using the data-driven distribution. The panels report the slope of the best linear fit of $\Delta I^{\textrm{peak}}_{i}$ versus $p_{M,i}$ as a function of $\beta$, for different values of $a$ and $b$, with fixed $p_M=0.2$. Each point corresponds to a simulation set used for fitting; error bars represent the 95\% confidence intervals. }\label{fig:Ipeak_slopes_all}
\end{figure}

Figure \ref{fig:DeltaAR_tiles_all}  plots the difference in the attack rate of the tiles between the data-driven distribution and the uniform distribution of non-compliant individuals.
Figures \ref{fig:slopes_AR_tiles_all} and \ref{fig:Ipeak_slopes_all} show the linear fit coefficient of the difference in attack rate and the difference in peak time of the tiles, respectively.
From these figures, it is clear that the observed behaviour  is consistent across the different cities, with only minor variations in the values of the fit coefficients.

\section{Effects of contact graph randomization}

\begin{figure}
	\centering
	\includegraphics[width=0.325\linewidth]{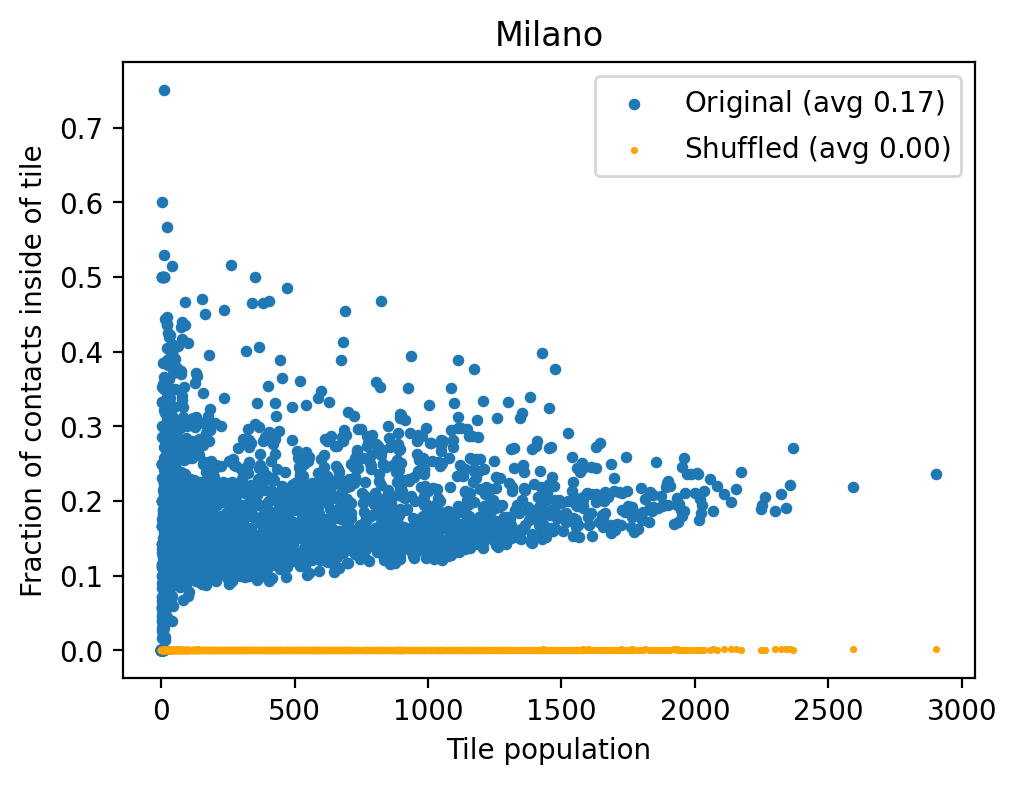}
	\includegraphics[width=0.325\linewidth]{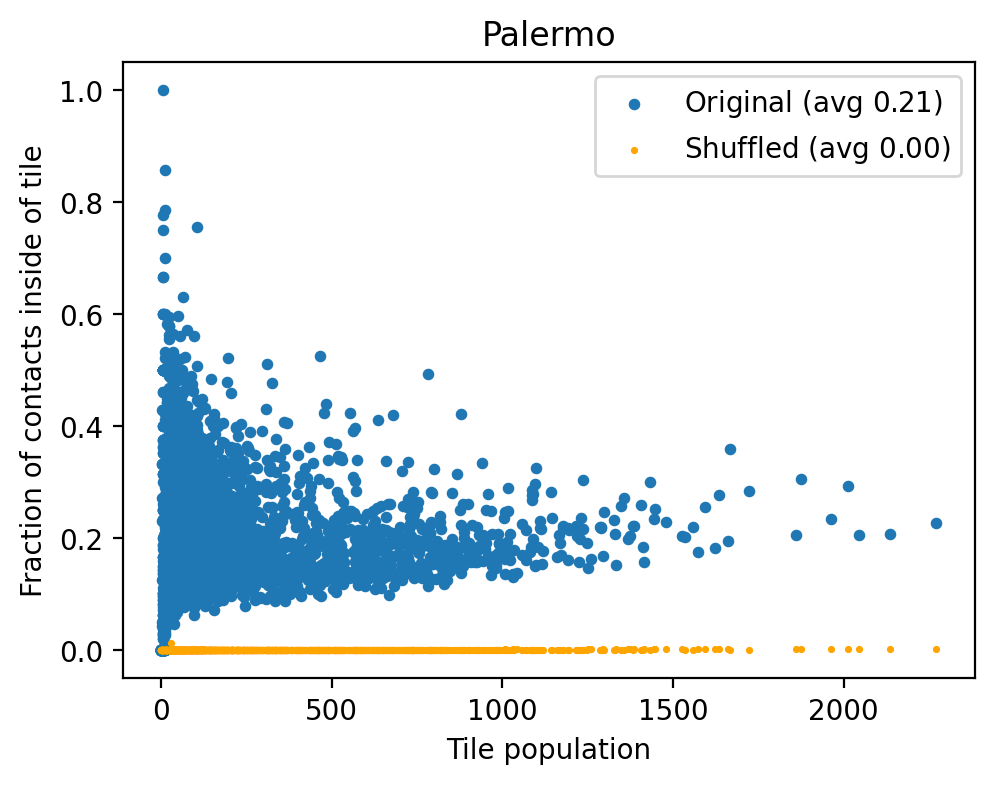}
	\caption{The scatter plots show the fraction of contacts that individuals in each tile have with individuals in the same tile, as a function of the tile population. The original city contact graphs contain a large number of intra-tile contacts, which are not retained by the shuffled graph.}
	\label{fig:diff_contacts}
\end{figure}

\begin{figure}
	\begin{centering}
    \includegraphics[width=0.7\linewidth]{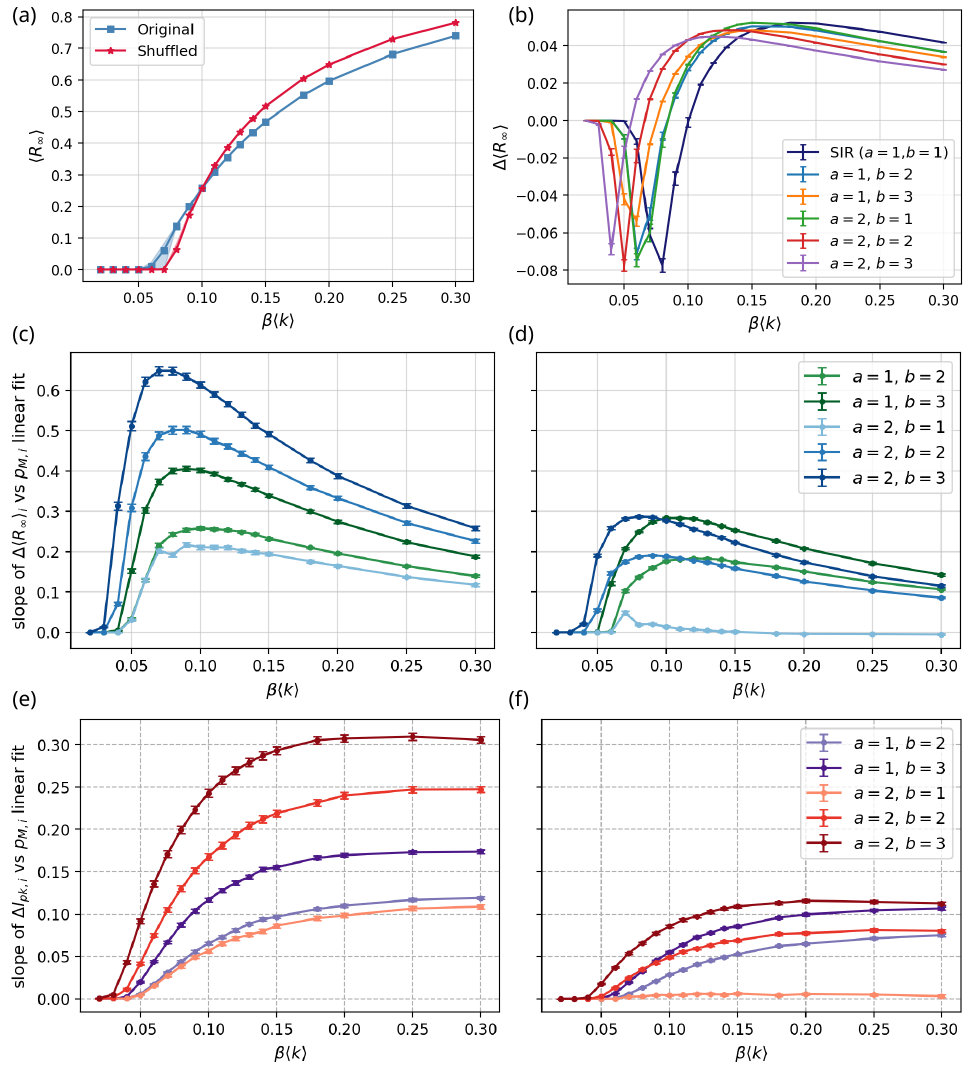}
    \par\end{centering}
	\caption{Comparison of results using original and shuffled contact graphs in the city of Palermo. 
Panel (a): Attack Rate (AR) of the SIR model for both graphs. 
Panel (b): Difference in AR (shuffled graph minus original graph) for the HeSIR and SIR models across various values of \(a, b\). 
Panels (c-d): Comparison of the slopes of the linear fit of \(\Delta \left< R_{\infty} \right>_i\) versus \(p_{M,i}\) as a function of \(\beta\), for different \(a, b\) values in the HeSIR model with data-driven distribution. On the left, the slopes for the original graph, on the right the slopes for the shuffled graph.
Panels (e-f): Comparison of the slopes of the linear fit of $\Delta I_{pk,i}$ versus $p_{M,i}$, for different \(a, b\) values in the HeSIR model with data-driven distribution, analogously as panel (c).}\label{fig:confmod_pa}
\end{figure}

The scatter plots in Figure \ref{fig:diff_contacts} show the proportion of contacts within each tile that involve individuals who both belong to that tile. It is clear from this result that household cliques are destroyed by randomisation, with almost all contacts in the shuffled contact networks being with people from other tiles.
Figure \ref{fig:confmod_pa} shows the epidemic effects of randomising the contact graph for the city of Palermo. These results are qualitatively similar to those in Milan, demonstrating that randomisation of the contact graph significantly impacts the epidemic outcomes observed.

